\pdfoutput=1
\documentclass[aps,prl,reprint,groupedaddress]{revtex4-1}

\usepackage{graphicx}
\usepackage{dcolumn}
\usepackage{bm}
\usepackage{physics}
\usepackage{amssymb}
\usepackage{natbib}

\begin{document}


\title{Borosilicones and viscoelastic silicone rubbers: network liquids and network solids}


\author{Louis A. Bloomfield}
\email{lab3e@virginia.edu}
\affiliation{Department of Physics, University of Virginia, Charlottesville, VA 22904}


\date{\today}

\begin{abstract}
Borosilicones such as bouncing putty (e.g., Silly Putty) have been known for more than 70 years, but the origins of their peculiar behaviors have remained a mystery. In this work, experiment and theory are used to show that borosilicones are network liquids---dynamic macromolecules that appear elastic on short timescales but exhibit flow on longer timescales. Each borosilicone is a vast covalent network of silicone polymer chains joined by trifunctional boron crosslinks. At any instant, the borosilicone is a highly-crosslinked elastic material. Because the boron crosslinks are temporary, however, the network evolves with time and the borosilicone exhibits liquid behavior.

A simple borosilicone, with chains coupled only by temporary crosslinks, exemplifies a classic transient network model and behaves as a simple (Lodge) elastic fluid. Its measured moduli and viscosities fit those predicted by both the transient network model and the Maxwell viscoelastic model: an elastic spring in series with a viscous dashpot. Those models lead to the same integer-order linear differential equations, which predict the exponential relaxation processes that are observed.

When some of the chains in a borosilicone are permanently crosslinked, however, the borosilicone no longer behaves as a simple elastic fluid. Though still a network liquid, this non-simple borosilicone exhibits slower-than-exponential relaxation processes that cannot arise from the integer-order linear differential equations representing finite arrangements of integer-order viscoelastic element (dashpots and springs). 

Infinite arrangements have no such limitations and the non-simple borosilicone can be modeled as a random assembly of infinitesimal viscoelastic bodies. Viscous bodies representing transient couplings occupy volume fraction $\beta$ and elastic bodies representing non-transient couplings occupy volume fraction $1-\beta$. An analytic study of this random assembly and a computational study of the analogous random network find behavior that is intermediate between viscous and elastic. Whereas a viscous material obeys differential equations of order 1 and an elastic material obeys differential equations of order 0, the random assembly obeys differential equations of fractional order $\beta$. The random assembly acts as a fractional-order viscoelastic element known as a spring-pot.

To eliminate the spring-pot's unphysical divergence as $\omega\to\infty$, a spring is placed in series with it. That pairing is the Fractional Maxwell viscoelastic model, the predictions of which agree well with the measured moduli and viscosities of non-simple borosilicones. Among those predictions are time-dependent viscosities that grow without limit and Mittag-Leffler-function relaxation processes.

When its concentration of permanent crosslinks exceeds the gelation threshold, a borosilicone becomes a viscoelastic silicone rubber (VSR). With a permanent network that spans the material coupled to a temporary network that also spans the material, the VSR is a network liquid piggybacking on a network solid. As a network solid, the VSR has an equilibrium shape to which it returns in the absence of external influences. As a network liquid with transient and non-transient couplings, the VSR has viscoelastic dynamics like those of a non-simple borosilicone. The Fractional Zener viscoelastic model, an elastic spring in parallel to the Fractional Maxwell model, accurately predicts the measured moduli of VSRs.

The temporary nature of boron crosslinks is due to exchange reactions in which -OH bearing molecules substitute for another endlessly in their covalent attachments to boron atoms. The energy barrier is small, so the mean lifetime of the temporary crosslinks is only modestly temperature-dependent. Because that mean lifetime gives rise to a borosilicone's only significant timescale, dominating its viscoelastic dynamics, a change in temperature merely shifts the borosilicone's timescale and the borosilicone is therefore thermo-rheologically simple.

\end{abstract}


\maketitle

\section{Introduction}

Borosilicones are a unique class of viscoelastic liquids and solids. Misunderstood and neglected scientifically for about seventy years, borosilicones are actually far more interesting than they originally appeared. In their simplest form, borosilicones are silicone polymer chains connected temporarily by boron crosslinks. In addition to having exceptional physical and mechanical properties, borosilicones exemplify some of the most basic and powerful theories in rheology, viscoelasticity, and random system, and offer unparalleled physical realizations of several basic viscoelastic models.

While the simplest borosilicones can be described using the models and mathematics of viscoelasticity texts, including differential equations of integer order, more complicated borosilicones require more sophisticated models and mathematics. These include differential equations of non-integer order, random network theory, and fractional viscoelastic model elements. Because the experimental measurements of borosilicones are best interpreted through those theoretical models and calculations, experiment and theory are woven together throughout this article.

Borosilicones were first prepared in the 1940s\cite{rochow1945} and their unusual properties were soon recognized with the discovery of silicone bouncing putty (SBP)\cite{wright1951}. A rudimentary borosilicone, SBP is a liquid that collapses slowly into a puddle under its own weight yet bounces beautifully as a ball dropped onto a hard surface. SBP eventually became the key ingredient in Silly Putty, a familiar children’s toy, and is sold in various forms to this day. Despite its unusual properties, however, SBP found little practical use outside the toy store and has remained mostly a curiosity for more than half a century.

Several explanations for SBP’s unusual behavior have been proposed over the years\cite{wick1960,freeman1962,mitrofanov1969,zatsepina1970}, with the current favorite being that SBP is composed of long polymer molecules that are entangled and linked by hydrogen bonds\cite{cross2012}. On short timescales, the hydrogen bonds and entanglements prevent flow and the SBP behaves as an elastic solid. On longer timescales, the hydrogen bonds can break, the molecules can disentangle, and the SBP can flow. SBP has been the textbook example of a thermoplastic polymer in the rubbery flow regime\cite{sperling2006}, a category it shares with the polymers found in chewing gum.

This paper will show that SBP is actually a Lodge Elastic Fluid, a viscoelastic fluid defined by a transient network theory first proposed by Green and Tobolsky in 1946\cite{green1946} and extended by Lodge in 1956\cite{lodge1956}. In their original transient network model, Green and Tobolosky explored the behavior of elastic strands networked together by chemical crosslinks that break and reform at a steady rate. Their work served to explain the slow stress relaxation observed in polysulfide rubbers at elevated temperatures.\cite{stern1946}

Although Green and Tobolsky's transient network model assumed chemical crosslinks, it found much wider application when Rouse\cite{rouse1953}, Zimm\cite{zimm1956}, and Lodge\cite{lodge1956} extended it to molten polymers and polymer solutions. Though not crosslinked in a chemical sense, the strands in molten polymers and polymer solutions interact with one another in ways that resemble transient crosslinks, notably through entanglements. Fluids exemplifying the transient network model are known as Lodge Elastic Fluids.

\section{Lodge Elastic Fluid}

Since it will be shown that silicone bouncing putty is a Lodge Elastic Fluid (LEF), a brief review of the transient network model is appropriate. That model builds on rubber elasticity theory, particularly the kinetic theory of elasticity\cite{shaw2005}, which finds that polymer strands of sufficient length and flexibility have a Gaussian distribution of end-to-end separations and behave as Hookean entropy springs.

The transient network (TN) model makes five key assumptions\cite{larson1988}:
\begin{enumerate}
\item The strands are Hookean entropy springs
\item The strands deform affinely until they release
\item Strands break with a constant probability per unit time, independent of the network deformation
\item Strands re-form as fast as they break
\item Strands re-form in configurations typical of equilibrium
\end{enumerate}

If the probability per unit time of a strand breaking and reforming is $1/\tau$, then the probability $P(t-t')$ of a strand surviving intact from time $t'$ to time $t$ satisfies the equation
\begin{equation}
\frac{d}{dt}P(t-t')=-\frac{1}{\tau}P(t-t').\label{eq:survivalDiffEq}
\end{equation}
Since $P(0)=1$, the solution to Eq.~(\ref{eq:survivalDiffEq}) is
\begin{equation}
P(t-t')=e^{-(t-t')/\tau}.\label{eq:survivalExpDecay}
\end{equation}
The time constant $\tau$ is thus the mean lifetime of the strands.

We can calculate the contribution $d {\tensor \sigma}_{t'}$ to the material's stress tensor ${\tensor \sigma}(t)$ due to a strand that re-formed at time $t'$ as the product of four quantities: the probability $dt'/\tau$ that the strand formed during the interval between $t'$ and $t'+dt'$, the probability $P(t-t')$ that this strand survived until time $t$, the material's modulus $G$, and the Finger tensor ${\tensor C}^{-1}(t,t')$  characterizing the deformation that occurred between time $t'$ and time $t$
\begin{equation}
d{\tensor \sigma}_{t'}=\frac{dt'}{\tau}e^{-(t-t')/\tau} G {\tensor {C}}^{-1}(t,t').\label{eq:dsigmaLEF}
\end{equation}

From rubber elasticity theory,  the modulus $G=\nu k T$, where $\nu$ is the number of strands per unit volume. Integrating Eq.~(\ref{eq:dsigmaLEF}) over all past times $t'$ gives the constitutive equation for the transient network model, also known as the Lodge equation:
\begin{equation}
{\tensor \sigma}(t) = G \int_{-\infty}^{t} \frac{1}{\tau}e^{-(t-t')/\tau}{\tensor C}^{-1}(t,t')dt'.\label{eq:constitutiveLEF}
\end{equation}

In general, a viscoelastic material's stress-strain relationship can be written\cite{shaw2012}
\begin{equation}
{\tensor \sigma}(t) = \int_{-\infty}^{t} \left[\frac{\partial}{\partial t'}G(t-t')\right]{\tensor C}^{-1}(t,t')dt',\label{eq:constitutiveGeneric}
\end{equation}
where $G(t-t')$ is the stress relaxation modulus. For the Lodge Elastic Fluid described by Eq. 
(\ref{eq:constitutiveLEF}), the stress relaxation modulus $G_{\textsc{lef}}(t-t')$ is
\begin{equation}
G_{\textsc{lef}}(t-t') = G e^{-(t-t')/\tau}.\label{eq:LEFStressRelaxationModulus}
\end{equation}
Thus, the stress in an LEF produced by a step in strain at time $t'=0$ decays away exponentially with characteristic time $\tau$, the mean lifetime of strands in the LEF.

\section{Simple Borosilicones}

There are a number techniques for incorporating boron in polyorganosiloxanes\cite{noll1968} and consequently for producing silicone bouncing putties\cite{mcgregor1947,dickmann1955,nitzsche1958,boot1965}. In the present work, simple borosilicones were prepared by reacting a silanol-terminated polydimethylsiloxane fluid OH-PDMS-OH with the liquid boron compound trimethyl borate B(OCH$_3$)$_3$. Rochow used a similar technique in 1940 to produce the first borosilicones\cite{rochow1945}, but did not immediately realize what had occurred. Reactions of silanol groups with alkoxy groups on boron were thought to have relatively low yields\cite{abel1959,noll1968}, however, Boot successfully produced SBPs in 1965 using such reactions\cite{boot1965} and it works well in practice.

The reaction between a silanol group (OH-Si) and a methoxy group on boron (RO-CH$_{3}$) is a condensation reaction
\begin{center}
	B(OCH$_{3}$)$_{3}$ + HO-PDMS-OH\\
	$\updownarrow$\\
	B(OCH$_{3}$)$_{2}$-O-PDMS-OH + CH$_{3}$OH
\end{center}
and results in a methanol molecule CH$_{3}$OH. This reaction is reversible, so that alcoholysis accompanies condensation, and it normally reaches an equilibrium in which both reactants and products are present.

To drive the reaction toward condensation and borosilicone formation, the methanol must be removed. Simply allowing the methanol to evaporate is sufficient, but vacuum drying propels the reaction to completion far more efficiently. The condensation reaction is so rapid, even at room temperature, that vacuum removal of the methanol can cause the borosilicone to form in seconds. Vacuum drying was used routinely throughout this work.

To produce borosilicones of the highest quality, only unblended, single-equilibration silanol-terminated polydimethylsiloxane (STPDMS) fluids were used in their preparation. Though commercial products rather than pure chemicals, these single-equilibration fluids had relatively narrow, single-peaked molecular weight (MW) distributions and minimal impurities.

The concentration of OH groups in each fluid could be estimated from its viscosity and used to determine the amount of trimethyl borate (TMB) needed to achieve stoichiometric saturation. When 3 mol of STPDMS react with 2 mol of TMB and the methanol is removed, all of the silanol groups of the STPDMS fluid (HO-PDMS-OH) are replaced by boron crosslinks (B-O-PDMS-O-B) and stoichiometric saturation is achieved.

Andisil OH 40 (AB Specialty Silicones) is a low-MW STPDMS fluid which nominally contains 3.5 wt\% OH groups. The product lot used in this work was certified as containing 3.99 wt\% OH, a value confirmed locally using Karl-Fischer titration\cite{bloomfield2012}. Adding 8.13 wt\% TMB to that fluid should result in stoichiometric saturation. 

To produce a saturated borosilicone, 8.13g TMB (Alfa-Aesar B20215) were added to 100.00g water-free Andisil OH 40 and the mixture was vacuum dried through two liquid-nitrogen-cooled Pyrex cold traps. Over a 30 minute period, 6.68g methanol were collect in those traps, 89 wt\% of the methanol expected if the condensation reaction went to completion. By that time, the borosilicone had become a brittle, translucent foam that made removing additional methanol difficult. The borosilicone was crushed and further vacuum dried, but the the residual methanol was not collected.

This simple borosilicone, designated SB40-8.13, is brittle, granular, and easily mistaken for a solid. It is actually an extraordinarily viscous liquid. When compressed in an arbor or hydraulic press, its granules slowly consolidate into a transparent mass. Moreover, SB40-8.13 flows under its own weight, but at such an imperceptible rate that days are required to observe any significant change in shape. [N.B. liquid flow is universal among simple borosilicones---despite hundreds of experiments in which many different STPDMS fluids were combined with various amounts of TMB and then fully dried, no solid simple borosilicones were ever observed.]

That SB40-8.13 remains liquid, despite being at or near stoichiometric saturation, defies the predictions of Flory-Stockmayer theory\cite{flory1941a,flory1941b,stockmayer1944}. According to that theory, a linear polymer (e.g. STPDMS) crosslinked by a trifunctional branching unit (e.g. boron) reaches the gelation threshold at only 50\% of stoichiometric saturation. At 100\% of stoichiometric saturation, it should be a network solid---a gigantic macromolecule in which a single covalently-bonded network extends throughout the entire material. With nearly all of its PDMS chains connected by boron crosslinks, SB40-8.13 should be a network solid and unable to flow.

The answer to this puzzle lies in the boron crosslinks themselves. \textit{Boron crosslinks are not permanent, they are temporary.} Unlike conventional crosslinks, which rarely break in normal circumstance, boron crosslinks  detach and re-attach on timescales measured in seconds or less. With its temporary crosslinks, SB40-8.13 is not a network solid at all, it is a \textit{network liquid}---a gigantic macromolecule in which a single covalently-bonded network extends throughout the material, but with covalent bonds that detach and re-attach frequently so that the network can evolve in topology and geometry.

\begin{figure}
\includegraphics{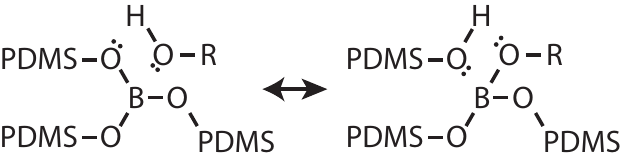}%
\caption{A proposed mechanism for the detachment and re-attachment of boron crosslinks. When an -OH group on a molecule aligns with a =BO- crosslink, the lone electron pairs on the two oxygen atoms facilitate a rearrangement of bonds. The previously unattached molecule crosslinks to the boron and a previously crosslinked ligand is released with a new -OH group. In this substitution, HO-R can be either a small molecule (e.g., water, an alcohol, or a carboxylic acid) or a silanol-terminated PDMS chain.\label{fig:exchangeReaction}}
\end{figure}

The mechanism by which boron crosslinks detach and re-attached almost certainly involves ligand substitution rather than actual bond breaking. In this substitution process, a =BO- moiety on a boron crosslink aligns with an -OH group on another molecule and, facilitated by their lone electron pairs, the two oxygen atoms trade bonding partners (Fig. \ref{fig:exchangeReaction}). The boron crosslink is left with a new ligand. That substitution process is similar to the four-center mechanism proposed for the rapid redistribution of alkoxy groups on boron observed in boron esters\cite{heyes1968}.

As evidence for such ligand substitution, it will be shown that the rate at which boron crosslinks detach and re-attach is extremely sensitive to the concentration of certain -OH groups, particularly -COOH groups, in a borosilicone. Adding just 1 ppm of iso-stearic acid (C$_{17}$H$_{35}$COOH) to a borosilicone significantly increases this rate. That -COOH groups boost the substitution rate more than other -OH groups probably stems from the second oxygen offering more possibilities for proper alignment and partner exchange. It will also be shown that the activation energy for boron crosslink detachment and re-attach is far too small to involve the breaking a covalent bond.

\subsection{Stress Relaxation Modulus}

A borosilicone's stress relaxation modulus $G(t)$ is measured by subjecting a small cylinder of that borosilicone to sudden compression. The 9.53mm-diameter cylinder is compressed from 6.35mm tall to 4.98mm tall in 10ms and held at that height while a load cell records the compressive force as a function of time. Dividing the compressive force by the cylinder's cross sectional area yields the compressive stress $\sigma_{c}(t)$, from which the stress relaxation modulus $G(t)$ can be calculated. Effects due to cylinder barreling, Coulomb friction\cite{williams2008}, and load-cell deformation are neglected.

When a cylinder's flat surfaces are perpendicular to axis 2 and compressed along axis 1, its total compressive stress $\sigma_{c}(t) = \sigma_{11}(t)-\sigma_{22}(t)$.\cite{shaw2012} Using Eq. (\ref{eq:constitutiveGeneric}),
\begin{equation}
\sigma_{c}(t) = \int_{-\infty}^{t}\left[\frac{\partial}{\partial t'}G(t-t')\right]\left(C_{11}^{-1}(t,t')-C_{22}^{-1}(t,t')\right)dt'.\label{eq:compressiveStress}
\end{equation}
For a compression step at $t'=0$, the Finger tensor elements are
\begin{eqnarray}
C^{-1}_{11}(t,t') - C^{-1}_{22}(t,t') & = & \lambda^{2}-\frac{1}{\lambda}\hspace{0.2cm}[t'<0]\nonumber\\
                                      & = & 0, \hspace{0.2cm} [t'\ge 0]\label{eq:compressionStepFinger}
\end{eqnarray}   
where $\lambda$ is the final height divided by initial height. Combining Eqs. (\ref{eq:compressionStepFinger}) and (\ref{eq:compressiveStress}) gives
\begin{eqnarray}
\sigma_{c}(t) & = & \int_{-\infty}^{0}\left[\frac{\partial}
	{\partial t'}G(t-t')\right]\left(\lambda^{2}-\frac{1}{\lambda}\right)dt'\nonumber\\
	      & = & -\left(\lambda^{2}-\frac{1}{\lambda}\right) G(t),
\end{eqnarray}
so that
\begin{equation}
G(t)=-\sigma_{c}(t)\left(\frac{\lambda}{1-\lambda^{3}}\right).\label{eq:compressionSRM}
\end{equation}
\begin{figure}
	\includegraphics{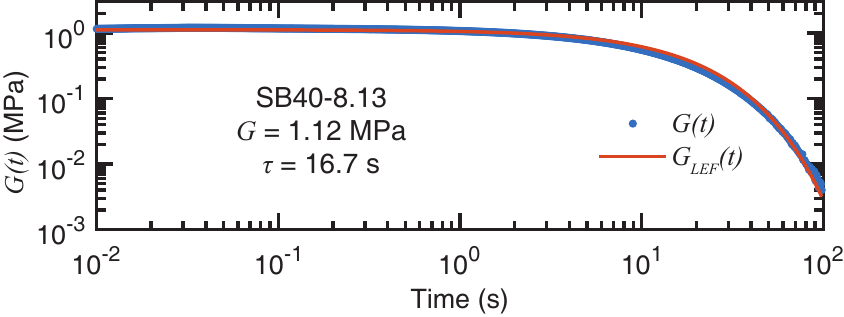}
	\caption{Stress relaxation modulus $G(t)$ of SB40-8.13, fit by the stress relaxation modulus $G_{\textsc{lef}}(t)$ of an LEF. Modulus $G$ and characteristic time $\tau$ are the fit's two parameters. The SB40-8.13 was in equilibrium with laboratory air.\label{fig:SB40-8p13StressRelaxationModulus}}
\end{figure}

Figure \ref{fig:SB40-8p13StressRelaxationModulus} shows SB40-8.13's stress relaxation modulus $G(t)$, obtained using the sudden compression technique. Also shown is a fit to those data by $G_{\textsc{lef}}(t)$, the stress relaxation modulus of an LEF from Eq. (\ref{eq:LEFStressRelaxationModulus}). The quality of the fit supports the hypothesis that SB40-8.13 is an LEF. It also provides values for the modulus $G$ and the characteristic time $\tau$. Repeated measurement yielded an experimental value of $G = 1.135 \pm 0.048$ MPa. Here and throughout this work, reported uncertainties are 2 standard deviations (95\% confidence).

For comparison, Andisil OH 40 was also cured with a conventional trifunctional crosslinker, methyltriethoxysilane (MTEOS). Using the technique of Mark\cite{mark1979}, 1.0 wt\% tin 2-ethylhexanoate catalyst were mixed into Andsil OH 40, followed by 10.66 wt\% MTEOS (stoichiometric saturation), and the mixture was cured in vacuum at 20 $^\circ$C for 10 days.

The cured silicone rubber was studied in the same compression apparatus used for the borosilicones. It was found to have modulus $G = 0.907\pm0.030$ MPa and no significant stress relaxation. That modulus is slightly less than the modulus of SB40-8.13, evidence that SB40-8.13 is a covalently-bonded network liquid. Rubber elasticity theory predicts $G\approx 3.0$ MPa.

After vacuum drying, SB40-8.13 has a characteristic time $\tau$ of about 40 s. $\tau$ is extremely sensitive to certain -OH groups, however, so it decreases as the borosilicone absorbs moisture from the air. When SB40-8.13 was allowed to equilibrate with laboratory air (20 $^\circ$C, 50 \% relative humidity), its $\tau$ decreased to about 17 s. 

\subsection{Shear Viscosity}

The shear viscosity $\eta(t)$ of a viscoelastic fluid is the ratio of shear stress $\sigma_{\textsc{s}}(t)$ to shear strain rate $\dot{\gamma}$,
\begin{equation}
\eta(t) = \frac{\sigma_{\textsc{s}}(t)}{\dot{\gamma}},\label{eq:defTDShearViscosity}
\end{equation}
for constant-rate shear strain that begins at time $t=0$. $\eta(t)$ is time-dependent because a viscoelastic fluid's shear stress $\sigma_{\textsc{s}}(t)$ depends on its shear strain at all earlier times and thus on $t$, the time since the start of shearing.

A borosilicone's shear viscosity $\eta(t)$ is measured by subjecting a 33mm-diameter by 4.0mm-thick disk to steady shear starting at time $t=0$, while a load cell records the shear force as a function of time. Dividing that shear force by the disk's cross sectional area and thickness gives the shear stress $\sigma_{\textsc{s}}(t)$, which is then divided by the shear strain rate $\dot{\gamma}$ to obtain the shear viscosity $\eta(t)$. Load cell deformation during the measurements is often significant enough to require compensation in order to maintain shear rates that are constant and accurate.

The time-dependent shear viscosity $\eta_{\textsc{lef}}(t)$ of an LEF can be calculated for comparison, beginning with a general viscoelastic fluid disk. When that disk's flat surfaces are perpendicular to axis 2 and sheared along axis 1, its shear stress $\sigma_{\textsc{s}}(t) = \sigma_{21}(t)$\cite{shaw2012} and Eq. (\ref{eq:constitutiveGeneric}) becomes
\begin{equation}
\sigma_{\textsc{s}}(t) = \int_{-\infty}^{t}\left[\frac{\partial}{\partial t'}G(t-t')\right]C_{21}^{-1}(t,t')dt'.\label{eq:shearStress}
\end{equation}
For shear strain at constant rate $\dot{\gamma}$ beginning at $t'=0$,
\begin{eqnarray}
C^{-1}_{21}(t,t') & = & \dot{\gamma}t\hspace{0.2cm}[t'<0]\nonumber\\
& = & \dot{\gamma}(t-t'). \hspace{0.2cm} [t'\ge 0]\label{eq:shearStepFinger}
\end{eqnarray}   
Combining Eqs. (\ref{eq:shearStress}) and (\ref{eq:shearStepFinger}) gives
\begin{eqnarray}
\sigma_{\textsc{s}}(t) & = & \int_{-\infty}^{t}\left[\frac{\partial}
					{\partial t'}G(t-t')\right]\dot{\gamma}t dt'\nonumber\\
              &   & -\int_{0}^{t}\left[\frac{\partial}
              {\partial t'}G(t-t')\right]\dot{\gamma}t' dt'\nonumber\\
			  & = & \dot{\gamma}\int_{0}^{t}G(t')dt',\label{eq:genericTDShearStress}
\end{eqnarray}
where it has been assumed that $tG(t) \to 0$ as $t\to\infty$.
Combining Eqs. (\ref{eq:defTDShearViscosity}) and (\ref{eq:genericTDShearStress}) gives
\begin{equation}
\eta(t) = \int_{0}^{t}G(t')dt'.\label{eq:genericTDShearViscosity}
\end{equation}

For an LEF, $G_{\textsc{lef}}(t)$ is given by Eq. (\ref{eq:LEFStressRelaxationModulus}) and the time-dependent shear viscosity $\eta_{\textsc{lef}}(t)$ is
\begin{eqnarray}
\eta_{\textsc{lef}}(t) & = & G\int_{0}^{t}e^{-t'/\tau}dt'\nonumber\\
        & = & G\tau(1-e^{-t/\tau}).\label{eq:LEFTDShearViscosity}
\end{eqnarray} 
The shear viscosity of an LEF thus starts at zero and approaches its long-time limit $\eta_{\textsc{lef}}=G\tau$ exponentially with characteristic time $\tau$. Since $\eta_{\textsc{lef}}$ is independent of the shear rate $\dot{\gamma}$, an LEF exhibits neither shear thinning nor shear thickening.

\begin{figure}
	\includegraphics{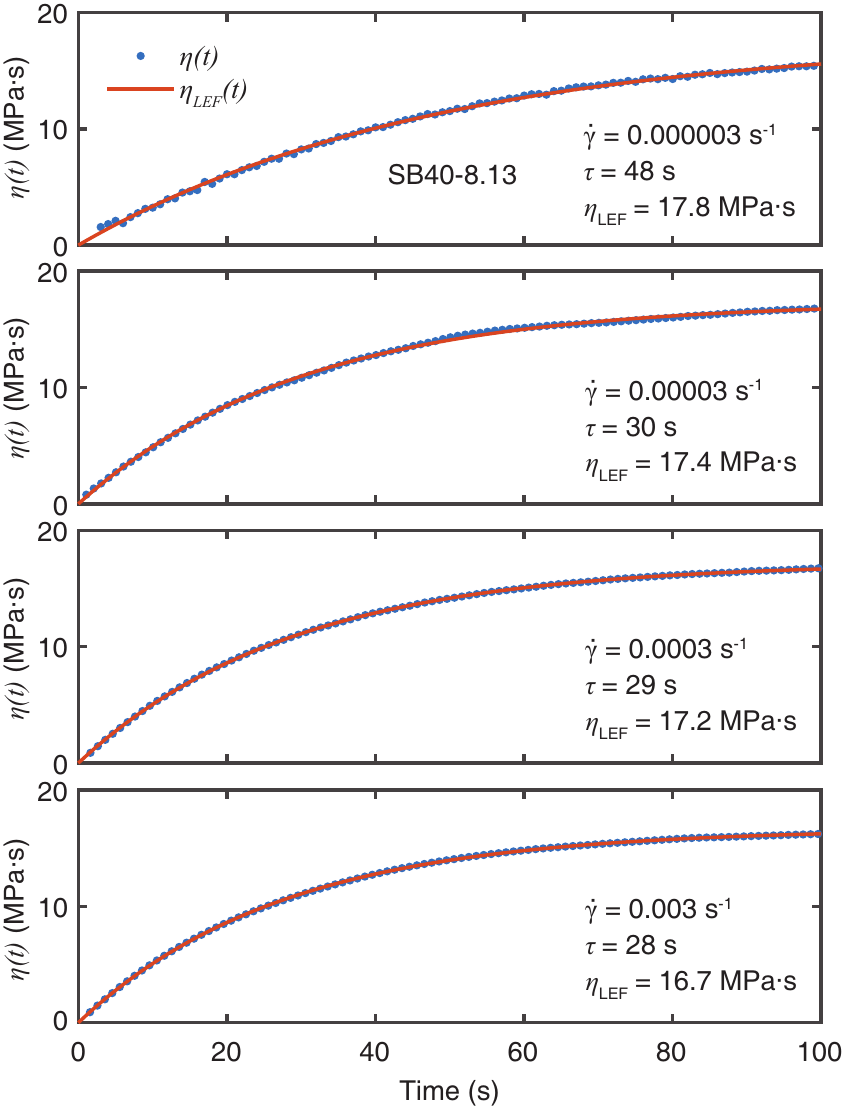}
	\caption{Shear viscosity $\eta(t)$ of SB40-8.13 at four different shear rates $\dot{\gamma}$. Each measurement is fit by the shear viscosity $\eta_{\textsc{lef}}(t)$ of an LEF. Limiting viscosity $\eta_{\textsc{lef}}$ and characteristic time $\tau$ are the fit's two parameters. The SB40-8.13 was drier than laboratory air. \label{fig:SB40-8p13ShearViscosity}}
\end{figure}

Measurements of the shear viscosity of SB40-8.13 at 4 different shear rates $\dot{\gamma}$ are shown in Fig. \ref{fig:SB40-8p13ShearViscosity}, along with fits by $\eta_{\textsc{lef}}(t)$. The fits are excellent, further evidence that SB40-8.13 is an LEF, and they provide values for the long-time shear viscosity $\eta_{\textsc{lef}}$ and the characteristic time $\tau$. $\eta_{\textsc{lef}}$ decreases by only 6\% as the shear rate increases by three orders of magnitude, nearly achieving the LEF prediction of no shear thinning at all.

The characteristic time $\tau$ decreases with increasing shear rate, particularly between $\dot{\gamma}=0.000003 s^{-1}$ and $\dot{\gamma}=0.00003 s^{-1}$. This effect is largely an instrumental artifact because, at the lowest shearing rates, the shearing motion is so slow that elastic deformation of the ``rigid'' instrument becomes significant. Although compensation is made for the increasing load cell deformation early in each measurement, some instrumental deformations remain uncompensated and affect $\tau$ in the slowest shear measurements. 

The modulus $G$ of SB40-8.13 obtained by dividing $\eta_{\textsc{lef}}$ by $\tau$ for many shear viscosity measurements is $G=0.584 \pm 0.141$ MPa. This $G$ is somewhat less than that obtained using the sudden compression technique. This difference is probably due to sample edge effects, barreling, and friction\cite{williams2008}, and to imperfect equipment. Improving absolute accuracy will be a focus of future studies.

\subsection{Elongation Viscosity}

The elongation viscosity $\eta_{\textsc{e}}(t)$ of a viscoelastic fluid is the ratio of elongation stress $\sigma_{\textsc{e}}(t)$ to elongation strain rate $\dot{\epsilon}$,
\begin{equation}
\eta_{\textsc{e}}(t) = \frac{\sigma_{\textsc{e}}(t)}{\dot{\epsilon}},\label{eq:defTDElongationViscosity}
\end{equation}
for constant-rate elongation strain that began at time $t=0$. $\eta_{\textsc{e}}(t)$ is time-dependent because a viscoelastic fluid's elongation stress $\sigma_{\textsc{e}}$ depends on its elongation strain at all earlier times and thus on $t$, the time since the start of elongation.

A borosilicone's elongation viscosity $\eta_{\textsc{e}}(t)$ is measured by stretching a rectangular beam so that its length increases exponentially in time, as required for constant-rate elongation. The beam is initially 6.54 mm x 6.54 mm x 14.22 mm long ($L_{0}$), and mounted in a stretching apparatus with the help of dog-bone ends. Starting at time $t=0$, this beam is stretched exponentially so that its length at $t>0$ is $L(t) = L_{0}e^{\dot{\epsilon} t}$. The exponential stretching continues for a specified time unless the beam breaks or the apparatus reaches its mechanical limits.

The time-dependent elongation viscosity $\eta_{\textsc{e,lef}}(t)$ of an LEF can be calculated for comparison, beginning with a general viscoelastic beam. When that beam is elongated along axis 1, its elongation stress $\sigma_{\textsc{e}}(t) = \sigma_{11}-\sigma_{22}$ and Eq. (\ref{eq:constitutiveGeneric}) becomes
\begin{equation}
\sigma_{\textsc{e}} = \int_{-\infty}^{t} \left[\frac{\partial}{\partial t'}G(t-t')\right]\left[C^{-1}_{11}(t,t') - C^{-1}_{22}(t,t')\right]dt'.\label{eq:elongationStress}
\end{equation}
For constant elongation strain rate $\dot{\epsilon}$, beginning at $t'=0$,
\begin{eqnarray}
C^{-1}_{11}(t,t') - C^{-1}_{22}(t,t') & = & \nonumber\\
& & e^{2\dot{\epsilon}t} - e^{-\dot{\epsilon}t}\hspace{0.2cm}[t'< 0]\label{eq:elongationStepFinger}\\
& & e^{2\dot{\epsilon}(t-t')} - e^{-\dot{\epsilon}(t-t')}\hspace{0.2cm}[t'\ge 0]\nonumber
\end{eqnarray}  
Combining Eqs. (\ref{eq:elongationStress}) and (\ref{eq:elongationStepFinger}) gives
\begin{eqnarray}
\sigma_{\textsc{e}}(t) & = &\int_{-\infty}^{0} \left[\frac{\partial}{\partial t'}G(t-t')\right] \left[e^{2\dot{\epsilon}t} - e^{-\dot{\epsilon}t}\right]dt'\nonumber\\
& & + \int_{0}^{t} \left[\frac{\partial}{\partial t'}G(t-t')\right] \left[e^{2\dot{\epsilon}(t-t')} - e^{-\dot{\epsilon}(t-t')}\right]dt'\nonumber\\
& = & 2\dot{\epsilon}\int_{0}^{t}G(t')e^{2\dot{\epsilon}t'}dt' + \dot{\epsilon}\int_{0}^{t}G(t')e^{-\dot{\epsilon}t'}dt',\label{eq:genericTDElongationStress}
\end{eqnarray}
where it has been assumed that $\left(e^{2\dot{\epsilon}t} - e^{-\dot{\epsilon}t}\right)G(t) \to 0$ as $t \to \infty$. 
Combining Eqs. (\ref{eq:genericTDElongationStress}) and (\ref{eq:defTDElongationViscosity}) gives
\begin{equation}
\eta_{\textsc{e}}(t)= 2\int_{0}^{t}G(t')e^{2\dot{\epsilon}t'}dt' + \int_{0}^{t}G(t')e^{-\dot{\epsilon}t'}dt'.\label{eq:genericTDElongationViscosity}
\end{equation}

For an LEF, $G_{\textsc{lef}}(t)$ is given by Eq. (\ref{eq:LEFStressRelaxationModulus}) and the time-dependent elongation viscosity $\eta_{\textsc{e,lef}}(t)$ is
\begin{eqnarray}
\eta_{\textsc{e,lef}}(t) & = & 2G\int_{0}^{t}e^{-t/\tau}e^{2\dot{\epsilon}t'}dt' + G\int_{0}^{t}e^{-t/\tau}e^{-\dot{\epsilon}t'}dt'\nonumber\\
 & = & \frac{2G\tau}{1-2\dot{\epsilon}\tau}(1-e^{-(1-2\dot{\epsilon}\tau)t/\tau})\nonumber\\
 &   & +\frac{G\tau}{1+\dot{\epsilon}\tau}(1-e^{-(1+\dot{\epsilon}\tau)t/\tau}).\label{eq:LEFTDElongationViscosity}
\end{eqnarray}

When $2\dot{\epsilon}\tau < 1$, $\eta_{\textsc{e,lef}}(t)$ asymptotically approaches a finite limit $\eta_{\textsc{e,lef}}$ at long times
\begin{equation}
\eta_{\textsc{e,lef}}=3G\tau\frac{1}{(1-2\dot{\epsilon}\tau)(1+\dot{\epsilon}\tau)}.\label{eq:LEFElongationViscosity}
\end{equation}
When $2\dot{\epsilon}\tau \ge 1$, however, $\eta_{\textsc{e,lef}}(t)$ increases without limit and does so exponentially in time. That effect is known as strain hardening\cite{larson1988} and leads inevitably to a broken beam if that beam is subjected to constant-rate elongation at $\dot{\epsilon} \ge 1/2\tau$. 

\begin{figure}
	\includegraphics{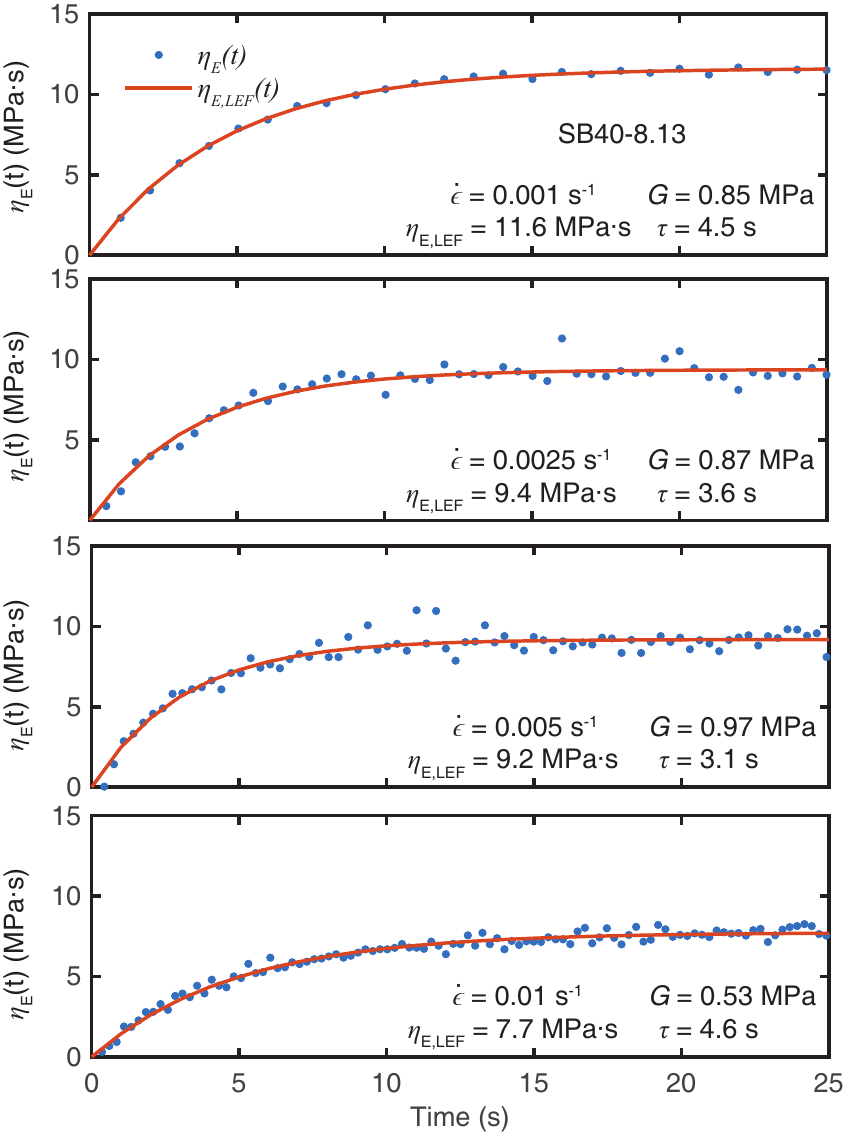}
	\caption{Elongation viscosity $\eta_{\textsc{e}}(t)$ of SB40-8.13 at four different elongation rates $\dot{\epsilon}$. Each measurement is fit by the elongation viscosity $\eta_{\textsc{e,lef}}(t)$ of an LEF. Modulus $G$ and characteristic time $\tau$ are the fit's two parameters. Limiting viscosity $\eta_{\textsc{e,lef}}$ is calculated from those parameters using Eq. (\ref{eq:LEFElongationViscosity}). To reduce its brittleness, the SB40-8.13 was moister than laboratory air.\label{fig:SB40-8p13ElongationViscosity}}
\end{figure}

Measurements of the elongation viscosity of SB40-8.13 at 4 different elongation rates $\dot{\epsilon}$ are shown in Fig. \ref{fig:SB40-8p13ElongationViscosity}, along with fits by $\eta_{\textsc{e,lef}}(t)$. The fits provide values for the modulus $G$ and the characteristic time $\tau$, from which SB40-8.13's limiting elongation viscosity $\eta_{\textsc{e,lef}}$ has been calculated using Eq. (\ref{eq:LEFElongationViscosity}). That elongation viscosity decreases slightly with increasing elongation rate. The values obtained for modulus $G$ range too widely to provide a meaningful average and confidence interval. 

Unreinforced borosilicones have poor tensile strengths and SB40-8.13 beams routinely break at elongation rates greater than about $0.015$ s$^{-1}$. It was thus not possible to exceed the strain hardening threshold $2\dot{\epsilon}\tau \ge 1$, which is $\dot{\epsilon} = 0.125$ s$^{-1}$ when $\tau=4$ s. Adding reinforcing fillers to this borosilicone would allow measurements in the strain hardening domain, however, those fillers introduce other important effects that will be discussed below.

\subsection{Complex Shear Modulus}

The complex shear modulus $G^{*}(\omega)$ of a viscoelastic material is the ratio of shear stress $\sigma_{\textsc{s}}(\omega,t)$ to shear strain $\epsilon(\omega,t)$ for steady oscillatory shear strain $\epsilon(t)=\epsilon_{0}e^{i\omega t}$,
\begin{equation}
G^{*}(\omega)=\frac{\sigma_{\textsc{s}}(\omega,t)}{\epsilon(\omega,t)}.\label{eq:defTDComplexShearModulus}
\end{equation}

A borosilicone's complex shear modulus $G^{*}(\omega)$ is measured using linear oscillatory shear strain. An aluminum septum is sandwiched between two disks of borosilicone, nominally 20mm dia x 1.4mm thick, and that assembly is sandwiched between two aluminum plates affixed to a load cell. An amplified waveform synthesizer and voice-coil linear actuator caused the septum to oscillate along its length relative to the fixed plates.

An optical system measures the oscillatory shear strain while the load cell measures the oscillatory shear force. Dividing the oscillatory shear force by the oscillatory shear strain gave the complex shear modulus times a calibration constant that depended on the exact geometries of the two sample disks.

For technical reasons, that calibration constant proved too difficult to control or measure accurately, so complex modulus values are reported in arbitrary units. At each angular frequency $\omega$, five or more oscillatory cycles were measured and analyzed separately in order to provide a value and a 95\% confidence interval for $G^{*}(\omega)$.

The complex shear modulus $G_{\textsc{lef}}^{*}(\omega)$ of an LEF can be calculated for comparison, starting with a general viscoelastic disk. When surfaces perpendicular to axis 2 are sheared along axis 1, shear stress $\sigma_{\textsc{s}}(t)$ is given by Eq. (\ref{eq:shearStress}). For steady oscillatory shear strain $\epsilon(t) = \epsilon_{0}e^{i\omega t}$,
\begin{equation}
C^{-1}_{21}(t,t')=\epsilon_{0}e^{i\omega t}-\epsilon_{0}e^{i\omega t'}.\label{eq:oscillatoryShearFinger}
\end{equation}
Combining Eqs. (\ref{eq:shearStress}) and (\ref{eq:oscillatoryShearFinger}) gives
\begin{eqnarray}
\sigma_{\textsc{s}}(t) & = & \int_{-\infty}^{t} \left[\frac{\partial}{\partial t'}G(t-t')\right]\left(\epsilon_{0}e^{i\omega t}-\epsilon_{0}e^{i\omega t'}\right)dt'\nonumber\\
              & = & i\omega \epsilon_{0}e^{i\omega t}\int_{0}^{\infty}G(t')e^{-i \omega t'}dt'\label{eq:genericTDOscillatoryShearStress}           
\end{eqnarray}
where it has been assumed that $G(t) \to 0$ as $t \to \infty$. Combining Eqs. (\ref{eq:defTDComplexShearModulus}) and (\ref{eq:genericTDOscillatoryShearStress}) gives
\begin{equation}
G^{*}(\omega) = i\omega \int_{0}^{\infty}G(t')e^{-i \omega t'}dt'.\label{eq:genericGStar}
\end{equation}

For an LEF, $G_{\textsc{lef}}(t)$ is given by Eq. (\ref{eq:LEFStressRelaxationModulus}) and the complex shear modulus is
\begin{eqnarray}
G_{\textsc{lef}}^{*}(\omega) & = & i\omega G\int_{0}^{\infty}e^{-t'/\tau}e^{-i \omega t'}dt'\nonumber\\
              & = & G\frac{i\omega\tau}{1+i\omega\tau}.\label{eq:LEFGStar}
\end{eqnarray}

A complex modulus can be separated into its real and imaginary parts
\begin{equation}
G^{*}(\omega) = G'(\omega) + i G''(\omega),\label{eq:genericGStarPrimes}
\end{equation}
where $G'(\omega)$ is the storage modulus and $G''(\omega)$ is the loss modulus. For an LEF, those two parts are
\begin{eqnarray}
G_{\textsc{lef}}'(\omega) & = & G\frac{\omega^{2}\tau^{2}}{1+\omega^{2}\tau^{2}}\label{eq:LEFGStarPrime}  \\
G_{\textsc{lef}}''(\omega) & = & G\frac{\omega\tau}{1+\omega^{2}\tau^{2}}.\label{eq:LEFGStarPPrime} 
\end{eqnarray}

\begin{figure}
	\includegraphics{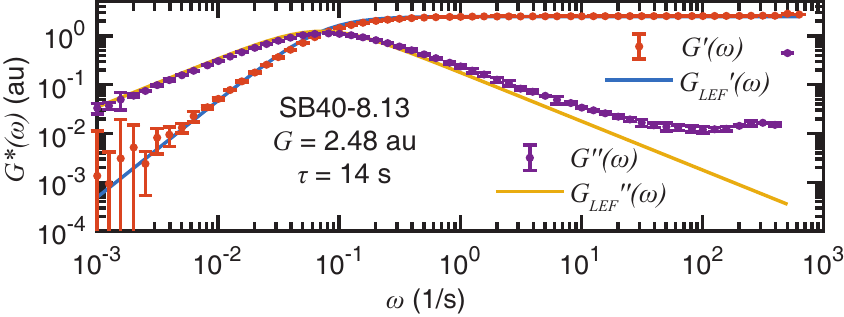}
	\caption{Complex shear modulus $G^{*}(\omega)= G'(\omega) + i G''(\omega)$ of SB40-8.13, fit by the complex shear modulus $G^{*}_{\textsc{lef}}(\omega)$ of an LEF. Modulus $G$ and characteristic time $\tau$ are the fit's two parameters. Lacking absolute calibration, $G$ is in arbitrary units. The SB40-8.13 was in equilibrium with laboratory air.\label{fig:SB40-8p13ComplexShearModulus}}
\end{figure}

A measurement of the complex shear modulus of SB40-8.13 is shown in Fig. \ref{fig:SB40-8p13ComplexShearModulus}, along with a fit by $G_{\textsc{lef}}^{*}(\omega)$. The fit provides values for the modulus $G$ and the characteristic time $\tau$, but $G$ is in arbitrary units due to the lack of absolute calibration.

At small $\omega$, the agreement between measurement and theory is excellent and SB40-8.13 is consistent with a Lodge Elastic Fluid. At large $\omega$, however, the measured loss modulus $G'(\omega)$ exceeds the predicted $G_{\textsc{lef}}'(\omega)$, suggesting that the simple borosilicone has fast dynamical processes that are not included of the transient network model.

Instrumental resonances sometimes interfere with measurements at specific large values of $\omega$. In Fig. \ref{fig:SB40-8p13ComplexShearModulus}, an instrumental resonance near $\omega = 628$ s$^{-1}$ disturbed the data for that $\omega$. 

\subsection{Maximizing the Modulus}

A simple borosilicone need not be at stoichiometric saturation to have a substantial modulus. Because boron crosslinks are dynamic, a borosilicone's modulus depends more on the statistics of its crosslinks than it does on how close the borosilicone is to stoichiometric saturation. Since boron crosslinks that coordinate only one or two PDMS chains contribute little to the modulus, maximizing the time-average concentration of boron crosslinks coordinating three PDMS chains should maximize the modulus.

Lacking a detailed study of the exchange reactions occurring at equilibrium, the moduli of these network liquids had to be maximized by trial and error. Borosilicones with different boron concentrations were prepared from the same lot of Andisil OH 40 fluid and their viscosities and moduli measured.

Near the gelation threshold, the borosilicones are soft enough that their viscosities can be measured with a falling sphere viscometer (Table \ref{tab:OH40Viscosities}). For a trifunctional crosslinker such as boron, the gelation threshold is 50\% of stoichiometric saturation\cite{flory1941a,flory1941b}. Since the Andisil OH 40 saturates at 8.13 wt\% TMB, its gelation threshold is 4.06 wt\% TMB.

The borosilicones indeed show a dramatic increase in viscosity as the gelation threshold is exceed. This observation is consistent with the idea that borosilicones are network liquids and that their covalently-bonded molecular networks grow to macroscopic size once the gelation threshold is exceeded. 

\begin{table}
\caption{Viscosities of Andisil OH 40 borosilicones measured using a falling sphere viscometer. For each borosilicone, the specified amount of trimethyl borate (TMB) was added to Andisil OH 40 and the mixture was vacuum dried.\label{tab:OH40Viscosities}}
\begin{ruledtabular}
\begin{tabular}{lccc}
	\textrm{Name}&
	\textrm{TMB (wt\%)}&
	\textrm{Viscosity (Pa$\cdot$s)}&
	Character\\
\colrule
OH 40 & 0.00 & $3.9\cdot 10^{-2}$ & thin liquid\\
SB40-4.00 & 4.00 & $1.4\cdot 10^{0}$ & syrup\\
SB40-4.10 & 4.10 & $3.0\cdot 10^{3}$ & soft gum\\
SB40-4.20 & 4.20 & $1.3\cdot 10^{4}$ & firm gum\\
SB40-4.50 & 4.50 & $6.5\cdot 10^{5}$ & gel\\
SB40-7.13 & 7.13 & $2.8\cdot 10^{7}$\footnote{Measured using a linear shear viscometer} & ``solid''
\end{tabular}
\end{ruledtabular}
\end{table}

\begin{figure}
	\includegraphics{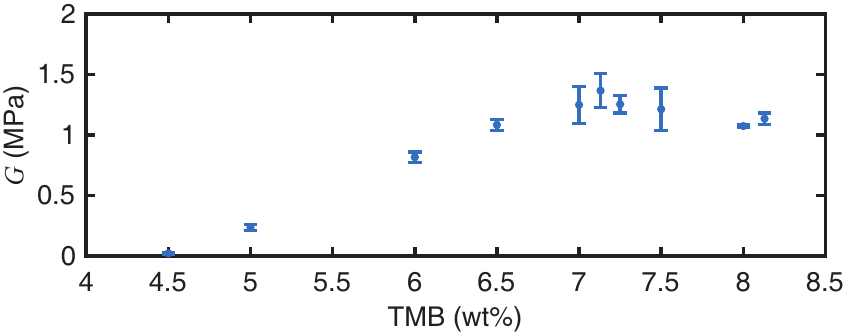}
	\caption{Moduli $G$ of Andisil OH borosilicones measured repeatedly using sudden compression. For each borosilicone, the specified amount of trimethyl borate (TMB) was added to Andisil OH 40 and the mixture was vacuum dried.\label{fig:OH40GvsTMB}}
\end{figure}

Further above the gelation threshold, the borosilicones are stiff enough that their moduli can be measured using the sudden compression technique (Fig. \ref{fig:OH40GvsTMB}). As the boron content increases, the modulus increases to a maximum at 7.13 wt\% TMB and then decreases until stoichiometric saturation is reached at 8.13 wt\% TMB (SB40-8.13).

The 7.13 wt\% borosilicone, designated SB40-7.13, was prepared by adding 14.26g TMB to 200.000g Andisil OH 40. Methanol released during the vacuum drying of SB40-7.13 was collected in liquid-nitrogen-cooled cold traps and totaled 92.3 wt\% of the methanol expected if the condensation reaction went to completion. The brittle borosilicone foam was then crushed and further dried to remove residual methanol. It has modulus $G = 1.366\pm0.140$ MPa. 

For comparison, a silicone rubber equivalent to SB40-7.13 was prepared using MTEOS and the curing procedure described above. It was found to have modulus $G = 0.877\pm 0.020$ MPa, significantly less than the modulus of its borosilicone equivalent.

That the borosilicone modulus peaks below stoichiometric saturation suggests that dynamic competition between the abundant boron crosslinks at saturation results in too many of those crosslinks being under-coordinated. A small reduction in boron crosslink concentration boosts their time-average coordination and results in a somewhat larger modulus.

Additionally, the saturated borosilicone SB40-8.13 becomes slightly cloudy when allowed to equilibrate with laboratory air. That cloudiness, almost certainly due to tiny crystallites of boric acid, is frequently observed in simple borosilicones with excessive boron concentrations. Dynamic competition between boron crosslinks and water molecules drives a small fraction of the boron out of solution as boric acid.

\begin{figure}
	\includegraphics{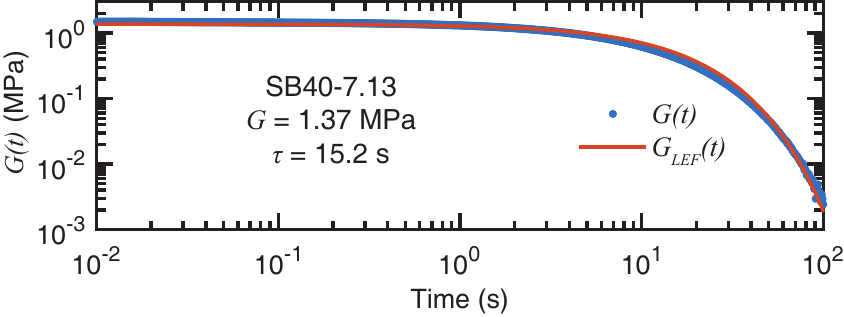}
	\caption{Stress relaxation modulus $G(t)$ of SB40-7.13, fit by the stress relaxation modulus $G_{\textsc{lef}}(t)$ of an LEF. Modulus $G$ and characteristic time $\tau$ are the fit's two parameters. The SB40-7.13 was in equilibrium with laboratory air.\label{fig:SB40-7p13StressRelaxationModulus}}
\end{figure}

A measurement of SB40-7.13's stress relaxation modulus $G(t)$ is shown in Fig. \ref{fig:SB40-7p13StressRelaxationModulus}, along with a fit by $G_{\textsc{lef}}(t)$. Many measurements yielded $G=1.366 \pm 0.140$ MPa, the largest modulus observed in the Andisil OH 40 borosilicones.

\begin{figure}
	\includegraphics{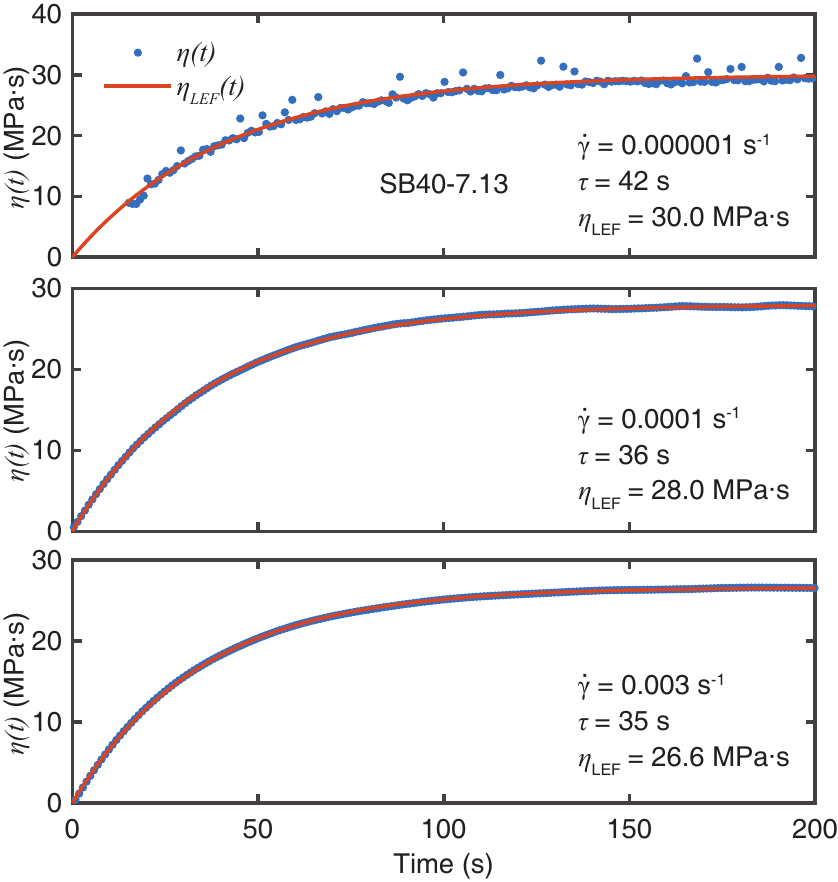}
	\caption{Shear viscosity $\eta(t)$ of SB40-7.13 at three different shear rates $\dot{\gamma}$. Each measurement is fit by the shear viscosity $\eta_{\textsc{lef}}(t)$ of an LEF. The SB40-7.13 was drier than laboratory air.\label{fig:SB40-7p13ShearViscosity}}
\end{figure}

Three measurements of SB40-7.13's shear viscosity $\eta(t)$ are shown in Fig. \ref{fig:SB40-7p13ShearViscosity}, each fit by $\eta_{\textsc{lef}}(t)$.
Slight shear thinning is observed as the shear rate increases. The modulus $G$ obtained by dividing $\eta_{\textsc{lef}}$ by $\tau$ for many shear viscosity measurements is $G=0.761 \pm 0.219$ MPa, somewhat less than the modulus obtained by the sudden compression technique.

\begin{figure}
	\includegraphics{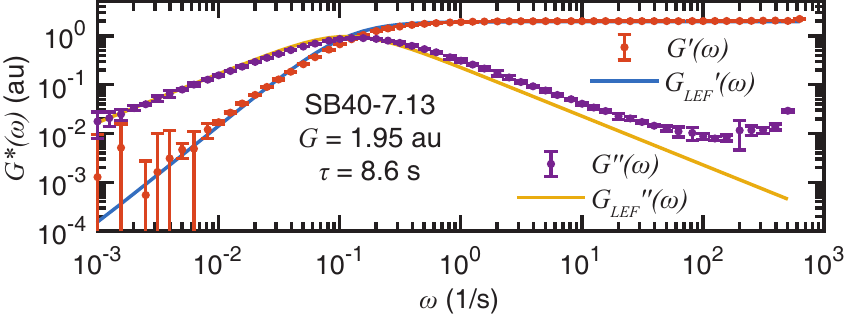}
	\caption{Complex shear modulus $G^{*}(\omega)= G'(\omega) + i G''(\omega)$ of SB40-7.13, fit by the complex shear modulus $G^{*}_{\textsc{lef}}(\omega)$ of an LEF. The SB40-7.13 was in equilibrium with laboratory air.\label{fig:SB40-7p13ComplexShearModulus}}
\end{figure}

A measurements of SB40-7.13's complex shear modulus $G^{*}(\omega)$ is shown in Fig. \ref{fig:SB40-7p13ComplexShearModulus}, along with a fit by $G^{*}_{\textsc{lef}}(\omega)$. The fit is excellent at small $\omega$, but the observed loss modulus $G''(\omega)$ modestly exceeds the prediction at large $\omega$.

\subsection{Other Simple Borosilicones}

\begin{table}
	\caption{Borosilicones were prepared from STPDMS fluids of six different viscosities. The average molecular weights (MW) are approximations based on viscosity. When the trimethyl borate (TMB) concentration shown was added to the STPDMS fluid and the mixture dried, the resulting simple borosilicone had the largest modulus $G$ observed for that fluid.\label{tab:STPDMS}}
	\begin{ruledtabular}
		\begin{tabular}{lrrcc}
			\textrm{Name}&
			\textrm{Viscosity}&
			\textrm{MW}&
			\textrm{TMB}&
			$G$ \textrm{}\\
			&
			\textrm{(cSt)}&
			\textrm{(Da)}&
			\textrm{(wt\%)}&
			\textrm{(MPa)}\\
			\colrule
			SB40-7.13	&40\footnote{Andisil OH fluid, AB Specialty Silicones\label{andisil}}
			& 850	& 7.13	& $1.366\pm0.140$\\
			SB70-2.60	& 70\footnote{Masil SFR fluids, Dystar\label{masil}}
			& 2700	& 2.60	& $0.685\pm0.036$\\
			SB100-2.20	&100\textsuperscript{\ref{masil}}
			& 4000	& 2.20	& $0.562\pm0.016$\\
			SB750-0.55	& 750\textsuperscript{\ref{masil}}
			& 18000	& 0.55	& $0.221\pm0.012$\\
			SB2000-0.33	& 2000\textsuperscript{\ref{masil}}
			& 36000	& 0.33	& $0.166\pm0.008$\\
			SB3500-0.28	& 3500\textsuperscript{\ref{masil}}
			& 42500	& 0.28	& $0.146\pm0.010$\\
		\end{tabular}
	\end{ruledtabular}
\end{table}

Simple borosilicones were also prepared from five other STPDMS fluids of increasing viscosity and average molecular weight (Table \ref{tab:STPDMS}). For each viscosity, borosilicones with slightly different TMB concentrations were prepared and studied to find the highest modulus. The stress relaxation moduli $G(t)$ and shear viscosities $\eta(t)$ of those peak-modulus simple borosilicones were then measured and fit by $G_{\textsc{lef}}(t)$ and $\eta_{\textsc{lef}}(t)$, respectively.

The fits agree well with the data, lending support to the idea that simple borosilicones are LEFs, and they yield values for the modulus $G$ and characteristic time $\tau$. $G$ decreases as the density of boron crosslinks decreases, consistent with rubber elasticity theory, and $\tau$ decreases with moisture content.

\begin{figure}
	\includegraphics{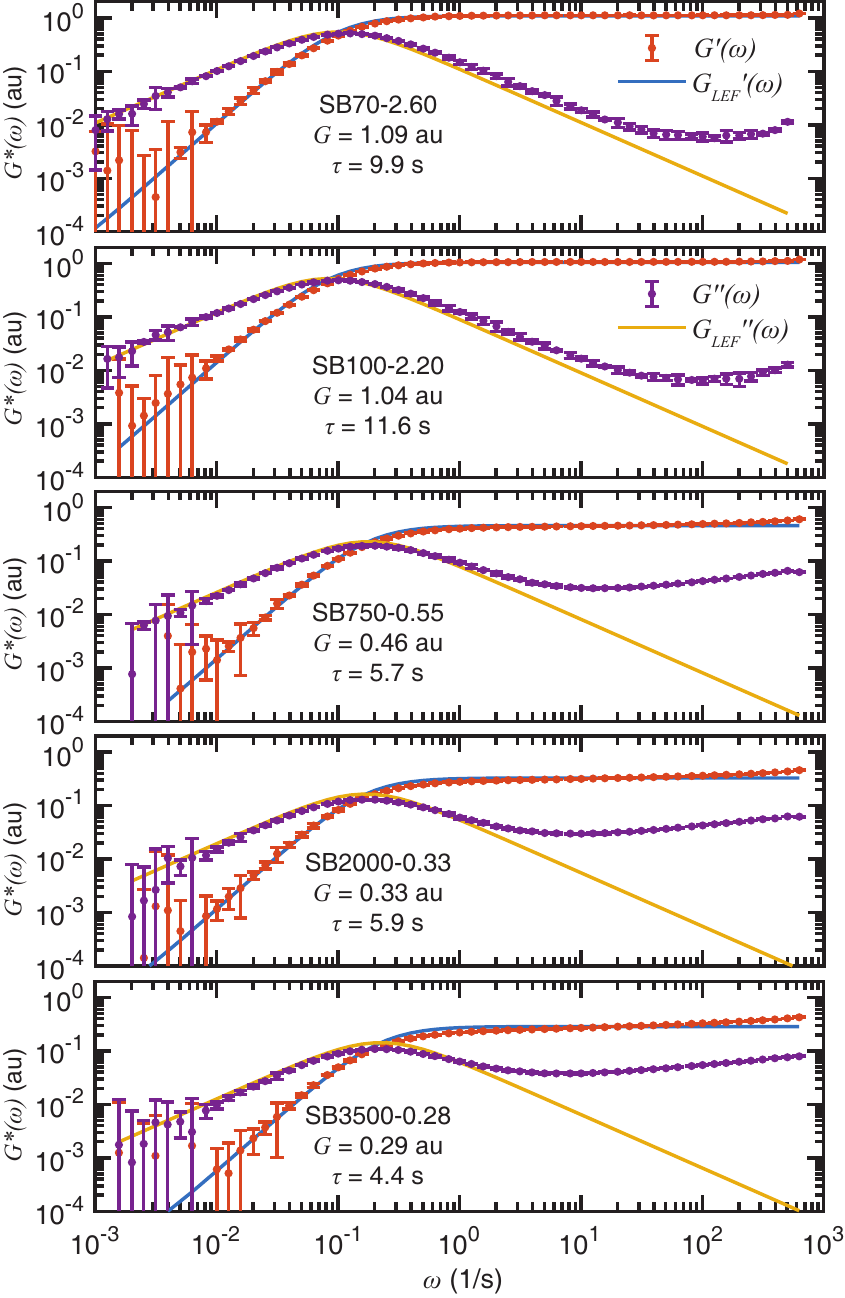}
	\caption{Complex shear moduli $G^{*}(\omega)= G'(\omega) + i G''(\omega)$ of simple borosilicones SB70-2.60 through SB3500-0.28, each fit by the complex shear modulus $G^{*}_{\textsc{lef}}(\omega)$ of an LEF.\label{fig:SBSeriesComplexShearModulus}}
\end{figure}

The measurements of complex shear modulus $G^{*}(\omega)$ are far more interesting (Fig. \ref{fig:SBSeriesComplexShearModulus}). At small $\omega$, each measured $G^{*}(\omega)$ agrees well with $G_{\textsc{lef}}^{*}(\omega)$ from Eq. (\ref{eq:LEFGStar}), consistent with the simple borosilicone being a Lodge Elastic Fluid. At large $\omega$, however, $G^{*}(\omega)$ and $G_{\textsc{lef}}^{*}(\omega)$ differ significantly.

For simple borosilicones made from lower viscosity fluids, SB40-7.13, SB70-2.60, and SB100-2.20, measurement and theory disagree only in the loss modulus $G''(\omega)$, with the simple borosilicones exhibiting modest excess loss. This excess is probably due to fast dissipative processes not included in the TN model.

For simple borosilicones made from higher viscosity fluids, SB750-0.55, SB2000-0.33, and SB3500-028, however, measurement and theory also disagree about the storage modulus $G''(\omega)$, with the simple borosilicones exhibiting modest excess storage. These excesses suggest that fast dynamical processes, both dissipative and non-dissipative, are present in long-chain borosilicones.

\section{Viscoelastic Models}
While Green and Tobolsky's transient network model and its Lodge Elastic Fluid predict the behaviors of simple borosilicones, they do less well with more complicated borosilicones. Once its PDMS chains are coupled to one another by more than just boron crosslinks, a borosilicone's network is not purely transient and its stress-strain behavior no longer obeys Eq. (\ref{eq:constitutiveLEF}), the Lodge Equation. 

Before examining more complicated borosilicones, a more complete theory is needed. To apply to borosilicones in which the strands are coupled to one another by more than just transient attachments, that extended theory must be able to describe a fluid whose network is intermediate between the transient network of an LEF and the permanent network of an elastic solid. 

The most expeditious path to that more complete theory involves viscoelastic modeling. Ordinary viscoelastic models can be found in any textbook on rheology or viscoelastic materials\cite{lakes2009,shaw2012} and will be reviewed here only briefly. They are generally built from combinations of two simple conceptual elements: springs and dashpots.

A spring element exhibits an elastic relationship between stress $\sigma(t)$ and strain $\epsilon(t)$:
\begin{equation}
\sigma(t) = G \epsilon(t),\label{eq:stressStrainSpring}
\end{equation}
where $G$ is the spring element's modulus. For steady oscillatory strain $\epsilon(\omega,t)=\epsilon_{0}e^{i\omega t}$, the stress is $\sigma(\omega,t)=G\epsilon_{0}e^{i\omega t}$ and the spring element's complex modulus $G_{\textsc{s}}^{*}(\omega)$ is
\begin{equation}
G_{\textsc{s}}^{*}(\omega)\equiv\frac{\sigma(\omega,t)}{\epsilon(\omega,t)} = G.
\end{equation}

A dashpot exhibits a viscous relationship between stress $\sigma(t)$ and strain $\epsilon(t)$:
\begin{equation}
\sigma(t) = \eta\frac{d}{dt}\epsilon(t).\label{eq:stressStrainDashpot}
\end{equation}
where $\eta$ is the dashpot's viscosity. For steady oscillatory strain $\epsilon(\omega,t)=\epsilon_{0}e^{i\omega t}$, the stress is $\sigma(\omega,t)=i\omega\eta\epsilon_{0}e^{i\omega t}$ and the dashpot element's complex modulus $G_{\textsc{d}}^{*}(\omega)$ is
\begin{equation}
G_{\textsc{d}}^{*}(\omega)\equiv \frac{\sigma(\omega,t)}{\epsilon(\omega,t)} = i\omega\eta
\end{equation}

\subsection{The Maxwell model}

\begin{figure}
	\includegraphics{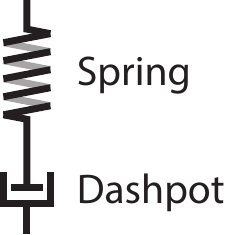}%
	\caption{The Maxwell model consists of a spring in series with a dashpot. \label{fig:MaxwellModel}}
\end{figure}

The Maxwell model for a viscoelastic fluid consists of only two elements: a spring in series with a dashpot (Fig. \ref{fig:MaxwellModel}). Its complex modulus $G_{\textsc{m}}^{*}(\omega)$ can be calculated from $G_{\textsc{s}}^{*}(\omega)$ and $G_{\textsc{d}}^{*}(\omega)$ using a series formula analogous to that for electrical conductance:
\begin{equation}
G_{\textrm{series}}^{*}(\omega) = \left(\frac{1}{G_{1}^{*}(\omega)}+\frac{1}{G_{2}^{*}(\omega)}\right)^{-1},\label{eq:series}
\end{equation}
where $G_{1}^{*}(\omega)$ and $G_{2}^{*}(\omega)$ characterize the individual elements. For the Maxwell model, \begin{eqnarray}
G_{\textsc{m}}^{*}(\omega) & = & \left(\frac{1}{G_{\textsc{s}}^{*}(\omega)}+\frac{1}{G_{\textsc{d}}^{*}(\omega)}\right)^{-1} \nonumber\\
& = & \left(\frac{1}{G}+\frac{1}{i\omega\eta}\right)^{-1} \nonumber \\
& = & G\frac{i\omega\tau}{1+i\omega\tau},\label{eq:MaxwellGStar}
\end{eqnarray}
where $\tau\equiv\eta/G$.

Since $G_{\textsc{m}}^{*}(\omega)$ has the same form as $G_{\textsc{lef}}^{*}(\omega)$, the complex shear modulus of an LEF in Eq. (\ref{eq:LEFGStar}), the Maxwell model perfectly characterizes the shear response of an LEF. It accurately predicts the shear responses of the simple borosilicones, at least for small $\omega$.

The Maxwell model's stress relaxation modulus $G_{\textsc{m}}(t)$ can be obtained from $G_{\textsc{m}}^{*}(\omega)$ via a simple transformation. For any fluid, $G(t)$ can be obtained from $G^{*}(\omega)$ by dividing Eq. (\ref{eq:genericGStar}) by $\omega$ and then operating on both sides with the one-sided Fourier transform $\int_{0}^{\infty}d\omega\sin(\omega t)$: 
\begin{equation}
G(t) = \frac{2}{\pi}\int_{0}^{\infty}d\omega\sin(\omega t)\frac{1}{\omega}\Re\left[G^{*}(\omega)\right].\label{eq:genericG}
\end{equation}
For the Maxwell model,
\begin{eqnarray}
G_{\textsc{m}}(t) & = & \frac{2}{\pi}\int_{0}^{\infty}d\omega\sin(\omega t)\frac{1}{\omega}G\frac{\omega^2\tau^2}{1+\omega^2\tau^2}\nonumber\\
& = & Ge^{-t/\tau}.\label{eq:MaxwellG}
\end{eqnarray}

Since $G_{\textsc{m}}(t)$ has the same form as $G_{\textsc{lef}}(t)$, the stress relaxation modulus of an LEF in Eq. (\ref{eq:LEFStressRelaxationModulus}), the Maxwell model also characterizes the compression responses of an LEF and the simple borosilicones. Although the Maxwell model is a one-dimensional concept, its constitutive equation is the same as that of an LEF, Eq. (\ref{eq:constitutiveLEF}).

\subsection{The Fractional Maxwell model}
	
The more complete theory, however, requires more than just springs and dashpots. It also requires a viscoelastic element that is intermediate between a spring and a dashpot, a fractional-order viscoelastic element known as a spring-pot\cite{koeller1984}.

Springs and dashpots are viscoelastic elements of integer order, meaning that their stress-strain relationships involve integer-order time derivatives. A spring's stress-strain relationship, Eq. (\ref{eq:stressStrainSpring}), can be written as
\begin{equation}
\sigma(t) = G\frac{d^{0}}{dt^{0}}\epsilon(t)
\end{equation}
and that of a dashpot, Eq. (\ref{eq:stressStrainDashpot}), can be written as 
\begin{equation}
\sigma(t) = G\tau\frac{d^{1}}{dt^{1}}\epsilon(t),
\end{equation}
where $\tau \equiv \eta/G$. Together, these relationships are
\begin{equation}
\sigma(t)=G\tau^{n}\frac{d^{n}}{dt^{n}}\epsilon(t),\label{eq:stressStrainN}
\end{equation}
where $n=1$ for a dashpot and $n=0$ for a spring. For steady oscillatory strain $\epsilon(\omega,t)=\epsilon_{0}e^{i\omega t}$, the stress is $\sigma(\omega,t)=\sigma_{0}e^{i\omega t}$ and the complex modulus $G_{n}^{*}(\omega)$ is
\begin{equation}
G_{n}^{*}(\omega) = G(i\omega\tau)^{n}.
\end{equation}

In ordinary calculus, the differentiation operator $D_{x}$ is taken only to integer powers $n$
\begin{equation}
D_{x}^{n} \equiv \frac{d^{n}}{dx^{n}}.
\end{equation}
In the branch of mathematical analysis known as fractional calculus, however, that differentiation operator can be taken to any real or even complex power $\alpha$
\begin{equation}
D_{x}^{\alpha} \equiv \frac{d^{\alpha}}{dx^{\alpha}}.
\end{equation}
Replacing the integer $n$ in Eq. (\ref{eq:stressStrainN}) with the real number $\beta$ results in a fractional-order stress-strain relationship
\begin{equation}
\sigma(t)=G\tau^{\beta}\frac{d^{\beta}}{dt^{\beta}}\epsilon(t)\label{eq:stressStrainBeta}
\end{equation}
that defines a spring-pot. For $\beta = 0$, the spring-pot reduces to a spring. For $\beta = 1$, it reduces to a dashpot. For $0 < \beta < 1$, the spring-pot is a fractional-order viscoelastic element with behaviors intermediate between those of a spring and those of a dashpot.

Any fractional differentiation operator $D_{x}^{\alpha}$ must reproduce ordinary differentiation when $\alpha$ is integer. When $\alpha$ is non-integer, however, there is some flexibility in $D_{x}^{\alpha}$. The operator used here is the Riemann-Liouville left-sided derivative\cite{podlubny1998,deOliveira2014},
\begin{equation}
^{\textrm{RL}}D_{a^{+}}^{\alpha}[f(x)] \equiv \frac{1}{\Gamma(n-\alpha)}\frac{d^{n}}{dx^{n}}\int_{a}^{x}(x-\xi)^{n-\alpha-1}f(\xi)d\xi,
\end{equation}
for the case in which $n = 0$ and $a = -\infty$. Denoted $_{-\infty}D_{x}^{\alpha}$, this specific fractional derivative operator is
\begin{equation}
_{-\infty}D_{x}^{\alpha}[f(x)] = \frac{1}{\Gamma(-\alpha)}\int_{-\infty}^{x}(x-\xi)^{-\alpha-1}f(\xi)d\xi.\label{eq:fractionalD}
\end{equation}

That operator can be used to find the spring-pot's complex modulus $G_{\beta}^{*}(\omega)$. For steady oscillatory strain $\epsilon(\omega,t)=\epsilon_{0}e^{i\omega t}$, Eq. (\ref{eq:stressStrainBeta}) is 
\begin{eqnarray}
\sigma(\omega,t) & = & G\epsilon_{0}\tau^{\beta} {_{-\infty}D_{t}^{\beta}}\left[e^{i\omega t}\right]\nonumber\\
& = & G\epsilon_{0}\tau^{\beta} \frac{1}{\Gamma(-\beta)}\int_{-\infty}^{t}(t-\xi)^{-\beta-1}e^{i\omega \xi}d\xi\nonumber\\
& = & G\epsilon_{0}\tau^{\beta}(i\omega)^{\beta}e^{i\omega t}.
\end{eqnarray}
The spring-pot's complex modulus $G_{\beta}^{*}(\omega)$ is
\begin{equation}
G_{\beta}^{*}(\omega)\equiv\frac{\sigma(\omega,t)}{\epsilon(\omega,t)}=G(i\omega\tau)^{\beta}.\label{eq:spring-potGStar}
\end{equation}
As it must, $G_{\beta}^{*}(\omega)$ reduces to $G_{\textsc{d}}^{*}(\omega)$ when $\beta = 1$ and $G_{\textsc{s}}^{*}(\omega)$ when $\beta = 0$.

\begin{figure}
	\includegraphics{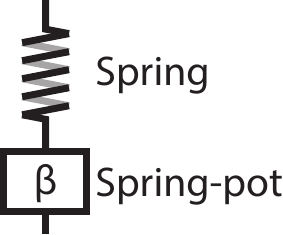}%
	\caption{The Fractional Maxwell model consists of a spring in series with a spring-pot. \label{fig:FractionalMaxwellModel}}
\end{figure}

When the Maxwell model's dashpot is replaced with a spring-pot (Fig. \ref{fig:FractionalMaxwellModel}), the result is the Fractional Maxwell model\cite{koeller1984,schiessel1995}. Consisting of a spring element in series with a spring-pot element, its complex modulus $G_{\textsc{fm}}^{*}(\omega)$ can be calculated using Eq. (\ref{eq:series}),
\begin{eqnarray}
G_{\textsc{fm}}^{*}(\omega) & = & \left(\frac{1}{G_{\textsc{s}}^{*}(\omega)}+\frac{1}{G_{\beta}^{*}(\omega)}\right)^{-1}\nonumber\\
& = & \left(\frac{1}{G}+\frac{1}{G(i\omega\tau)^{\beta}}\right)^{-1} \nonumber\\
& = & G\frac{(i\omega\tau)^{\beta}}{1+(i\omega\tau)^{\beta}}.\label{eq:FractionalMaxwell}
\end{eqnarray}
With its three parameters $G$, $\tau$, and $\beta$, the FM model includes all possible choices of spring and spring-pot.

The Fractional Maxwell model's stress relaxation modulus $G_{\textsc{fm}}(t)$ can be obtained from its complex modulus $G_{\textsc{fm}}^{*}(\omega)$ using Eq. (\ref{eq:genericG}), 
\begin{eqnarray}
G_{\textsc{fm}}(t) & = & \frac{2}{\pi}\int_{0}^{\infty}d\omega
\frac{\sin(\omega t)}{\omega}
\Re \left[\frac{G(i\omega\tau)^{\beta}}{1+(i\omega\tau)^{\beta}}\right]\nonumber\\
& = & \frac{G}{\pi}\int_{-\infty}^{\infty}d\omega\left[\frac{e^{i\omega t}-e^{-i\omega t}}{2i\omega}\right]\left[1-\frac{1}{1+(i\omega\tau)^{\beta}}\right]\nonumber\\
& = & G - \frac{G\tau^{-\beta}}{2\pi i}\int_{\gamma- i\infty}^{\gamma+ i\infty}e^{s t}ds\left[\frac{1}{s}\right]\frac{1}{\tau^{-\beta}+s^{\beta}}\nonumber\\
& & + \frac{G\tau^{-\beta}}{2\pi i}\int_{\gamma- i\infty}^{\gamma+ i\infty}e^{-s t}ds\left[\frac{1}{s}\right]\frac{1}{\tau^{-\beta}+s^{\beta}}.
\end{eqnarray}
where $s \equiv i\omega$ and $\gamma$ is a positive real infinitesimal. The two integrals are inverse Laplace transforms,
\begin{eqnarray}
G_{\textsc{fm}}(t) & = & G - G\tau^{-\beta}\mathcal{L}^{-1}\left\lbrace \frac{1/s}{a+s^{\beta}};t\right\rbrace\nonumber\\
 & &  + G\tau^{-\beta}\mathcal{L}^{-1}\left\lbrace \frac{1/s}{a+s^{\beta}};-t\right\rbrace,\label{eq:twoInverseLaplaceTransforms}
\end{eqnarray}
where $a = \tau^{-\beta}$. As shown in Ref. \cite{mathai2005}, Eq. (2.2.21),
\begin{equation}
\mathcal{L}^{-1}\left\lbrace \frac{1/s}{a+s^{\beta}};t\right\rbrace =
\int_{0}^{t}u(t')(t-t')^{\beta-1}E_{\beta,\beta}(-a(t-t')^{\beta})dt',
\end{equation}
where $u(x)$ is the unit step function. Because of the step function, the last term in Eq. (\ref{eq:twoInverseLaplaceTransforms}) is zero and
\begin{eqnarray}
G_{\textsc{fm}}(t) & = & G - G\tau^{-\beta}\int_{0}^{t}(t-t')^{\beta-1}E_{\beta,\beta}(-a(t-t')^{\beta})dt'\nonumber\\
& = & GE_{\beta}\left(-(t/\tau)^{\beta}\right),\label{eq:FMaxwellStressRelaxation}
\end{eqnarray}
where Ref. \cite{haubold2011}, Eq. (5.1) has been used. 

$E_{\alpha,\gamma}(z)$ and $E_{\alpha}(z)$ are the Mittag-Leffler functions,
\begin{eqnarray}
E_{\alpha}(z) & = & \sum_{k=0}^{\infty}\frac{z^{k}}{\Gamma(\alpha k + 1)}\\
E_{\alpha,\gamma}(z) & = & \sum_{k=0}^{\infty}\frac{z^{k}}{\Gamma(\alpha k + \gamma)},
\end{eqnarray}
generalized exponentials that reduce to ordinary exponentials when $\alpha=1$ and $\gamma=1$. Correspondingly, the Fractional Maxwell model's complex modulus $G_{\textsc{fm}}^{*}(\omega)$ and stress relaxation modulus $G_{\textsc{fm}}(t)$ reduce to those of the ordinary Maxwell model when $\beta=1$.

The constitutive equation for a material that exemplifies the Fractional Maxwell model can be obtained from $G_{\textsc{fm}}(t)$ by substituting its derivative\cite{haubold2011}
\begin{equation}
\frac{\partial}{\partial t'}G_{\textsc{fm}}(t-t') = \frac{G}{\tau}\left(\frac{t-t'}{\tau}\right)^{\beta-1} E_{\beta,\beta}\left(-\left(\frac{t-t'}{\tau}\right)^{\beta}\right)
\end{equation}
into Eq. (\ref{eq:constitutiveGeneric}), resulting in
\begin{eqnarray}
{\tensor \sigma}(t) & = & \int_{-\infty}^{t}  
\frac{G}{\tau}\left(\frac{t-t'}{\tau}\right)^{\beta-1}\nonumber\\
& & \hspace{1cm} E_{\beta,\beta}\left(-\left(\frac{t-t'}{\tau}\right)^{\beta}\right)
{\tensor C}^{-1}(t,t')dt'\label{eq:constitutiveFM}.
\end{eqnarray}
When $\beta=1$, this constitutive equation reduces to Eq. (\ref{eq:constitutiveLEF}), the Lodge Equation and the constitutive equation for the Maxwell model.

\subsection{Random Assembly Model}

The Fractional Maxwell (FM) model has been introduced because, when the PDMS chains of a borosilicone are coupled by more than just boron crosslinks, the borosilicone's network is no longer purely transient and the Lodge equation and Maxwell model no longer apply. Such borosilicones will be called ``non-simple'' because they have characteristics intermediate between those of  elastic fluids with their temporary networks and those of elastic solids with their permanent networks.

This intermediate territory is the domain of fractional-order viscoelastic models and fractional-order elements such as the spring-pot. Since the Maxwell model successfully predicts the behaviors of simple borosilicones, it is reasonable to suppose that the FM model, with a spring-pot replacing the Maxwell model's dashpot, will similarly predict the behaviors of non-simple borosilicones.   

To begin building a case for the FM model, consider the microscopic structure of a non-simple borosilicone. It has both transient and non-transient couplings between its chains and those interactions produce viscous and elastic behaviors, respectively\footnote{While transient couplings might be expected to produce elastic fluid behavior rather than viscous behavior, models based on that choice do not predict the behavior of non-simple borosilicones}. The material can be modeled as a randomly-distributed mixture of viscous regions with complex shear modulus $Gi\omega\tau$ and elastic regions with complex shear modulus $G$.

\begin{figure}
	\includegraphics{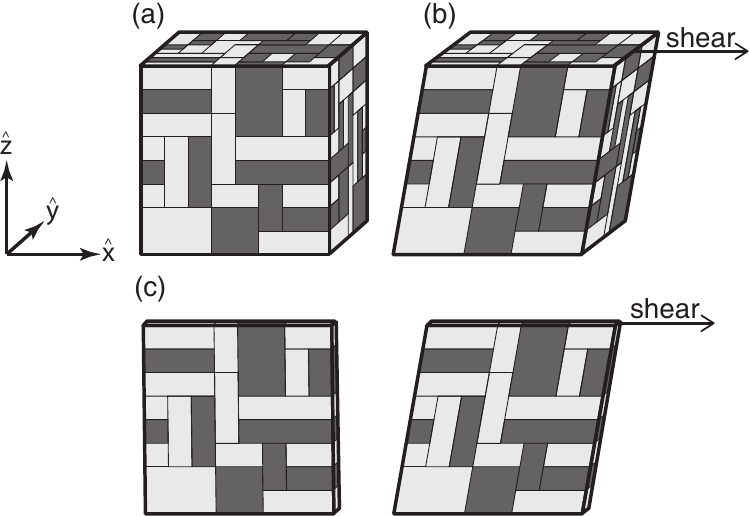}
	\caption{(a) A random assembly of upright rectangular cuboids (edges along $\hat{x}$, $\hat{y}$, and $\hat{z}$). (b) When surfaces perpendicular to $\hat{z}$ are sheared along $\hat{x}$, there are no interactions in the $\hat{y}$ direction and the random assembly can be reduced to (c) a thin slice parallel to the $xz$ plane.\label{fig:randomAssemblyShear}}
\end{figure}

Regardless of a region's shape, it can be deconstructed into a collection of upright rectangular cuboids (rectangular cuboids with edges along $\hat{\textrm{x}}$, $\hat{\textrm{y}}$, and $\hat{\textrm{z}}$). The entire material can thus be modeled as a random assembly of viscous and elastic cuboids (Fig. \ref{fig:randomAssemblyShear}a). A parameter $\beta$ establishes the volume fractions of viscous cuboids ($\beta$) and elastic cuboids ($1-\beta$).

Although the random assembly is three-dimensional, when a surface perpendicular to $\hat{\textrm{z}}$ is sheared in the $\hat{\textrm{x}}$ direction  (Fig. \ref{fig:randomAssemblyShear}b), its complex shear modulus $G_{\textsc{ra}}^{*}(\omega)$ depends only on its structure in the $xz$ plane. This observation assumes affine deformation and large-scale homogeneity of the model, so that there are no interactions in the $\hat{\textrm{y}}$ direction and a thin slice cut parallel to the xz plane  (Fig. \ref{fig:randomAssemblyShear}c) has the same complex shear modulus as the entire model. If the slice is cut thin enough, it will be uniform in the $\hat{y}$ direction and thus effectively two-dimensional. 

\begin{figure}
	\includegraphics{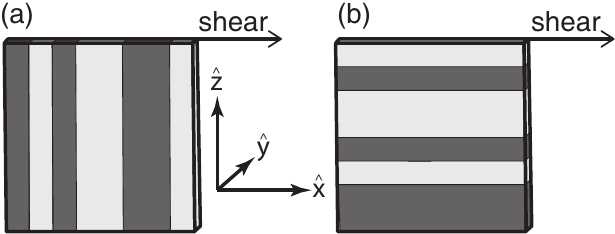}
	\caption{(a) When cuboids extend along $\hat{z}$, they act in parallel under shear. (c) When cuboids extend along $\hat{x}$, they act in series.\label{fig:parallelSeries}}
\end{figure}

If all the slice's cuboids extended in the $\hat{\textrm{z}}$ direction from surface to surface (Fig. \ref{fig:parallelSeries}a), then those cuboids would act in parallel and the slice's complex shear modulus would be
\begin{equation}
G^{*}_{\textrm{parallel}}(\omega)^{1} = \beta (Gi\omega\tau)^{1} + (1-\beta) (G)^{1}.
\end{equation}
If all of its cuboids extended in the $\hat{\textrm{x}}$ direction from surface to surface (Fig. \ref{fig:parallelSeries}b), then those cuboids would act in series and the slice's complex shear modulus would be
\begin{equation}
G^{*}_{\textrm{series}}(\omega)^{-1} = \beta (Gi\omega\tau)^{-1} + (1-\beta) (G)^{-1}.
\end{equation}
Writing the parallel and series formulas in this fashion allows them to be expressed together as
\begin{equation}
G^{*}_{\nu}(\omega)^{\nu} = \beta (Gi\omega\tau)^{\nu} + (1-\beta) (G)^{\nu},\label{eq:nuAssembly}
\end{equation}
where $\nu=1$ for parallel and $\nu=-1$ for series.

Of course, the slice's actual cuboid structure is far more complicated, with cuboids that are beside one another in the $\hat{\textrm{x}}$ direction acting in parallel and cuboids that are above one another in the $\hat{\textrm{z}}$ direction acting in series. Because the cuboids are arranged randomly, parallel and series connections are equally abundant. 

Following the mixing method of Lichtenecker\cite{lichtenecker1926}, $G_{\textsc{ra}}^{*}(\omega)$ can be obtained by evaluating Eq. (\ref{eq:nuAssembly}) in the limit $\nu\to 0$, where series and parallel connections are equally common. Expanding $z^{\nu}$ near $\nu=0$ as
\begin{equation}
z^{\nu}=1 + ln(z)\nu + \mathcal{O}(\nu^2),
\end{equation}
the limit of Eq. (\ref{eq:nuAssembly}) as $\nu\to 0$ is
\begin{equation}
ln(G_{\textsc{ra}}^{*}(\omega)) = \beta ln(Gi\omega\tau) + (1-\beta)ln(G).\label{eq:lnAssembly}
\end{equation}
Exponentiating both sides of Eq. (\ref{eq:lnAssembly}) gives
\begin{eqnarray}
G_{\textsc{ra}}^{*}(\omega) & = & (Gi\omega\tau)^{\beta} \cdot G^{1-\beta}\nonumber\\ 
& = & G(i\omega\tau)^{\beta}.\label{eq:randomAssemblyGStar}
\end{eqnarray}

$G_{\textsc{ra}}^{*}(\omega)$ has the same form as the complex modulus of a spring-pot $G_{\beta}^{*}(\omega)$, given in Eq. (\ref{eq:spring-potGStar}), so a random assembly of elastic and viscous cuboids acts as a spring-pot. Moreover, the parameter $\beta$ is both the fractional concentration of viscous cuboids in the random assembly and the fractional-order of the spring-pot's stress-strain relationship, Eq. (\ref{eq:stressStrainBeta}). These results support the idea that a non-simple borosilicone's mixture of transient and non-transient chain couplings can produce fractional-order viscoelastic behavior. 

\subsection{Random Network Model}

Most viscoelastic models are one-dimensional, so it would be nice to find a one-dimensional analog to the random assembly model. Fortunately, the random assembly model transforms easily into a one-dimensional random network model. This transformation begins with a thin slice of the random assembly, uniform in the $\hat{y}$ direction, and replaces each three-dimensional viscoelastic cuboid with a one-dimensional viscoelastic element. The complex modulus of that element is the cuboid's complex shear modulus multiplied by the cuboid's aspect ratio $a = \textrm{width}/\textrm{height}$. A viscous cuboid is thus replaced by a dashpot with complex modulus $aGi\omega\tau$ and an elastic cuboid is replaced by a spring with complex modulus $aG$.

\begin{figure}
	\includegraphics{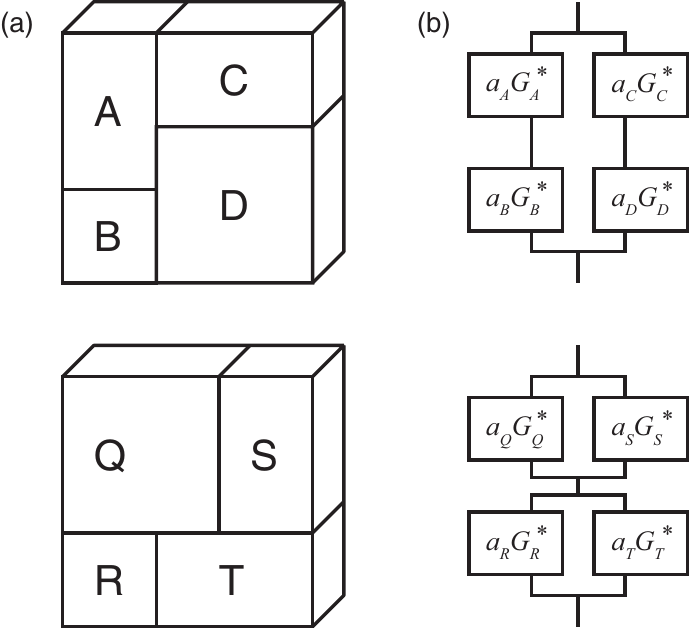}
	\caption{(a) Two 4-cuboid portions of random assembly are transformed into (b) two 4-element portions of random network. The modulus of an element is the modulus $G^{*}_{\textrm{cuboid}}$ of its cuboid's material times the aspect ratio $a_{\textrm{cuboid}}$ of its cuboid. Series pairs AB and CD combine in parallel. Parallel pairs QS and RT combine in series.\label{fig:RandomAssemblyToRandomNetwork}}
\end{figure}

The viscoelastic elements are then coupled together in series or parallel, using the slice of random assembly as a guide. Elements replacing cuboids that are above one another in the $\hat{z}$ direction couple in series, while elements replacing cuboids that are beside one another in the $\hat{x}$ direction couple in parallel. The two 4-cuboid portions of the random assembly shown in Fig. \ref{fig:RandomAssemblyToRandomNetwork}a are thus transformed into the two 4-element portions of the random network shown in Fig. \ref{fig:RandomAssemblyToRandomNetwork}b. Cuboids/elements A and B couple in series, C and D couple in series, and the coupled pairs then couple in parallel. Similarly, Q and S couple in parallel, R and T couple in parallel, and the coupled pairs then couple in series.

The overall random network is a vast hierarchical latticework of coupled subnetworks that are themselves coupled subnetworks. Despite its complexity, this random network can be drawn in two dimensions without its coupling lines crossing. The density of its elements is non-uniform, a consequence of the non-uniform number-density of the cuboids they replace. Since one element can always be divided into two half-elements in series or in parallel, however, the model could be expanded into an evenly-spaced square array of elements with appropriate couplings. Its structure would then look similar to the random networks of capacitors and resistors used to model complex conductivity and dielectric response in materials\cite{lichtenecker1926,truong1995,almond1999,wu2003,bowen2006}.

Since the random assembly behaves as a spring-pot, so does the corresponding random network. Despite being composed entirely of dashpots and springs, this random network acts as a fractional-order viscoelastic element. While fractional-order viscoelastic elements have been constructed previously out of integer-order elements\cite{schiessel1993,heymans1994,schiessel1995}, those constructions have involved highly ordered arrangements such as infinite ladders, trees, and fractals. For thermodynamic reasons, such orderly structures are unlikely to appear in real materials. Moreover, their fractional order $\beta$ is determined by network structure rather than a continuous parameter, usually limiting $\beta$ to a small number of discrete values.

In contrast, the random assembly and random network models discussed here use randomness and statistics to their advantage, so they should be easier to realize in real materials. Moreover, the fractional order $\beta$ that characterizes these models can easily take any value $1 \ge \beta \ge 0$.

\subsection{The Computational Random Network}

The Lichtenecker mixing method used to obtain the random assembly's complex shear modulus analytically is simple and elegant, but also rather mysterious. To verify its results and obtain further insight into these models, the random network model was studied computationally.

The computational approach begins by constructing a thin random assembly, using a recursive procedure to cut and recut a square region of the xz plane into billions of tiny rectangles. The inconsequential $\hat{y}$ dimension is omitted for simplicity. The rectangles are then randomly assigned viscous or elastic complex shear moduli, weighted statistically to ensure that the volume fraction of viscous rectangles is $\beta$ and of elastic rectangles is $1-\beta$.

Each rectangle is then replaced by a viscoelastic element with a complex modulus that is the rectangle's complex shear modulus multiplied by its aspect ratio. For computational purposes, that element's complex modulus is represented by its values at 200 discrete angular frequencies covering 20 orders of magnitude in $\omega$. For simplicity, $\tau \equiv 1$ s.

As the recursive procedure returns up through its many levels, its viscoelastic elements are coupled and recoupled to form a hierarchy of subnetworks that eventually become the entire random network. Any pair of subnetworks formed by cutting a rectangle along the $\hat{z}$ direction is coupled in parallel. Any pair formed by cutting a rectangle along the $\hat{x}$ direction is coupled in series. Each coupling is performed computationally on the two complex-value arrays representing its subnetworks, producing a single array that is returned to the next higher level. The final coupling yields a complex-value array representing the complex modulus of the entire random network.

\begin{figure}
	\includegraphics{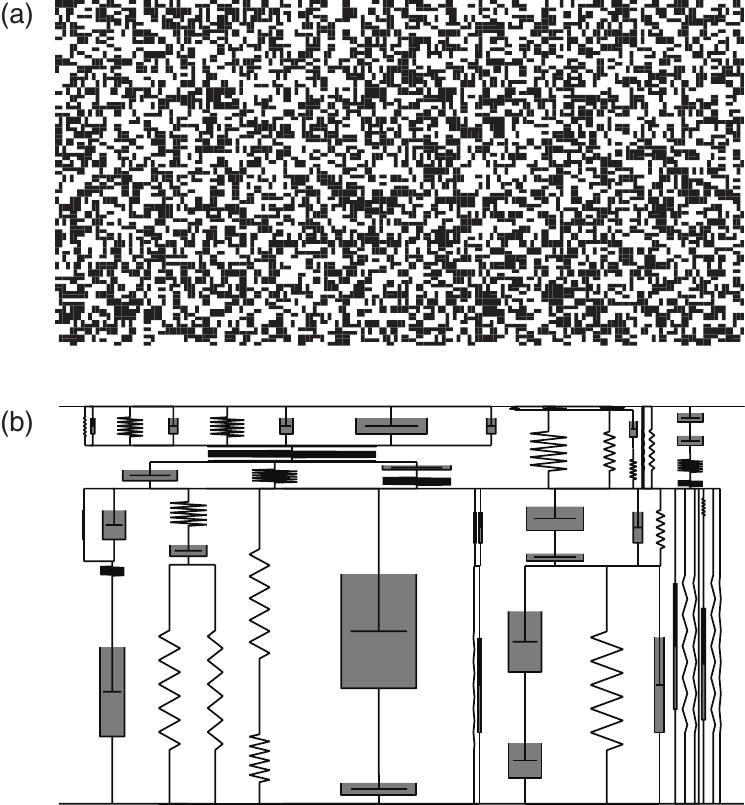}
	\caption{(a) A fraction of the rectangles in a random assembly with $\beta = 0.5$. Black represents viscous material and white represents elastic material. (b) A fraction of the dashpots and springs in a random network with $\beta = 0.5$.\label{fig:figRandomAssemblyNetwork}}
\end{figure}

A recursion depth of 36 levels gives clear, compelling results. Since each level of recursion divides a rectangle into two smaller rectangles, 36 levels of recursion produces a random assembly with $2^{36} = 68,719,476,736$ rectangles. The corresponding random network has 68,719,476,736 viscoelastic elements. Figure \ref{fig:figRandomAssemblyNetwork} displays (a) a small fraction ($1.3\cdot 10^{-7}$) of the random assembly with $\beta = 0.5$ and (b) an even smaller fraction ($8\cdot 10^{-10}$) of the random network.

\begin{figure}
	\includegraphics{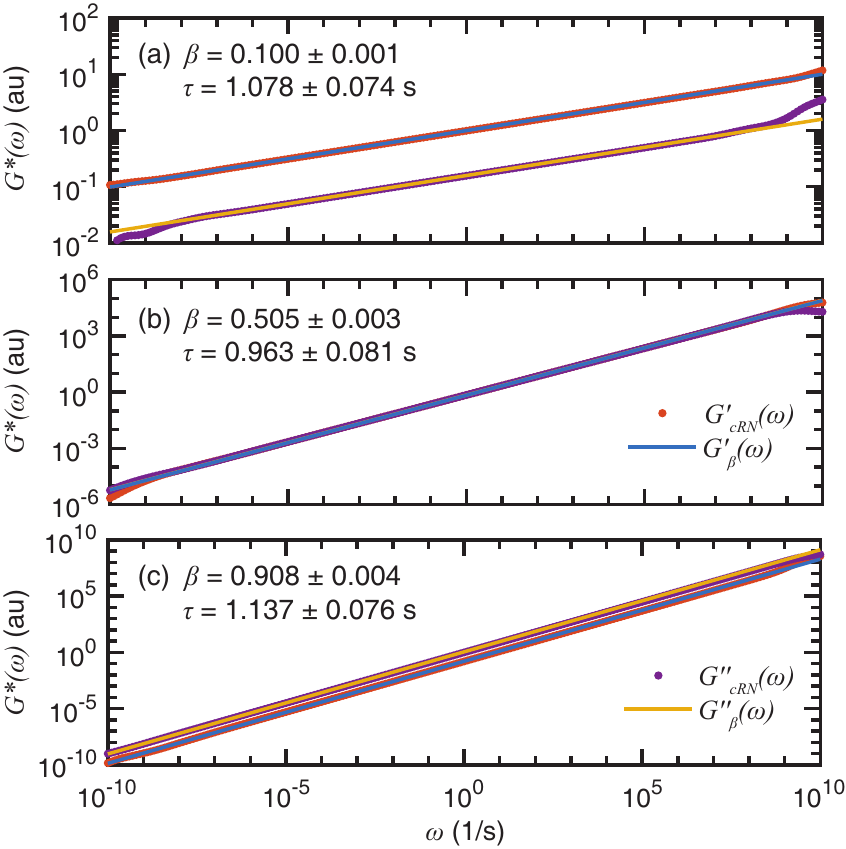}
	\caption{The computational random network model's complex modulus $G_{c{\textsc{rn}}}^{*}(\omega)$ for (a) $\beta = 0.1$, (b) $\beta = 0.5$, and (c) $\beta = 0.9$, with $\tau=1$ s and $G=1$.  Also shown are fits to the spring-pot's complex modulus $G_{\beta}^{*}(\omega)$, along with values of $\beta$ and $\tau$ obtained from those fits. \label{fig:randomNetworkComplexModulus}}
\end{figure}

Figure \ref{fig:randomNetworkComplexModulus} shows the computational random network's complex modulus $G_{c\textsc{rn}}^{*}(\omega)$ for (a) $\beta = 0.1$, (b) $\beta = 0.5$, and (c) $\beta = 0.9$, with its real and imaginary parts separated into its storage modulus $G_{c\textsc{rn}}'(\omega)$ and its loss modulus $G_{c\textsc{rn}}''(\omega)$, respectively. Also shown are fits by the spring-pot's complex modulus $G_{\beta}^{*}(\omega)$, again separated into storage and loss moduli,
\begin{eqnarray}
G_{\beta}'(\omega) & = & G(\omega\tau)^{\beta}\cos(\pi\beta/2)\\
G_{\beta}''(\omega) & = & G(\omega\tau)^{\beta}\sin(\pi\beta/2).
\end{eqnarray}
The fits are nearly perfect over 15 orders of magnitude in $\omega$ and the values of $\beta$ and $\tau$ obtained from those fits are almost identical to those used when constructing the computational random network.

It was observed that when constructing the random assembly, the choice of cut direction, whether along $\hat{x}$ or $\hat{z}$, had to be random. Any non-random sequence of cut directions (e.g., along $\hat{x}$ at even recursion levels, along $\hat{y}$ at odd levels) resulted in complex moduli that did not resemble that of a spring-pot. In contrast, the choice of cut amount did not have to be random. Whether a rectangle was cut into equal or random subrectangles had no significant effect on the overall complex modulus. The data in Fig. \ref{fig:randomNetworkComplexModulus} were produced by cutting rectangles into equal subrectangles.

\subsection{Spring-pot to Fractional Maxwell model}

Although the spring-pot arises naturally when both transient and non-transient couplings are included in the RA and RN models, the spring-pot's complex modulus diverges when $\omega\tau \to \infty$ for $\beta >0$. The spring-pot alone is therefore not a realistic viscoelastic model for non-simple borosilicones.

One way to avoid the divergence is to reconsider the materials used in the random assembly. Since a simple borosilicone exhibits Maxwell model behavior, perhaps the random assembly should contain a Maxwell material $G_{\textsc{m}}^{*}(\omega)$ rather than a viscous material. Making that substitution and applying the Lichtenecker mixing method to the modified random assembly yields complex shear modulus
\begin{equation}
G_{m\textsc{ra}}^{*}(\omega) = G\left(\frac{i\omega\tau}{1+i\omega\tau}\right)^\beta.
\end{equation}

$G_{m\textsc{ra}}^{*}(\omega)$ does indeed converge when $\omega\tau\to\infty$. Moreover, it reduces to $G_{\textsc{m}}^{*}(\omega)$ when $\beta=1$ and $G_{\textsc{s}}^{*}(\omega)$ when $\beta=0$. But despite these promising characteristics, $G_{m\textsc{ra}}^{*}(\omega)$ fails the most important test: it does not fit the measured complex shear moduli of non-simple borosilicones.

Another way to avoid the divergence is to recognize the finite compressibility of real liquids. A physically reasonable model for an ordinary liquid requires a compressible spring in series with a dashpot, resulting in the Maxwell model. Correspondingly, a physically reasonable model for a non-simple borosilicone requires a compressible spring in series with a spring-pot, resulting in the Fractional Maxwell model.

As desired, the FM model's complex modulus $G_{\textsc{fm}}^{*}(\omega)$, given in Eq. (\ref{eq:FractionalMaxwell}), converges when $\omega\tau\to\infty$. It also reduces to $G_{\textsc{m}}^{*}(\omega)$ when $\beta=1$ and, after first taking the limit as $\omega\tau\to\infty$, to $G_{\textsc{s}}^{*}(\omega)$ when $\beta=0$. More importantly, however, $G_{\textsc{fm}}^{*}(\omega)$ passes the critical test: it fits the measured complex shear moduli of non-simple borosilicones.

\section{Non-Simple Borosilicones}

Non-simple borosilicones have both transient and non-transient couplings between their PDMS chains. The transient couplings are temporary (boron) crosslinks, but the non-transient couplings can vary. The classic non-transient coupling is a permanent crosslink, so the quintessential non-simple borosilicone is a PDMS network with a mixture of temporary and permanent crosslinks. When those crosslinks are randomly distributed, the network is a physical realization of the RA and RN models.

One way to form those permanent crosslinks is through a condensation reaction with a silane crosslinker. Unfortunately, most silane crosslinkers require catalysts that remain in the borosilicone and affect its properties. Acetoxysilanes (=Si-CH$_{3}$COOH), however, will crosslink STPDMS fluids at elevated temperatures without catalysts. Non-simple borosilicones were therefore produced using vinyltriacetoxysilane (VTAS) as the permanent crosslinker.

To avoid the extreme stiffness and brittleness of borosilicones based on 40 cSt STPDMS (Andisil OH 40), non-simple borosilicones were instead based on 70 cSt STPDMS (Dystar Masil SFR 70). The permanent crosslinks were formed by adding VTAS to SFR 70 and heating the mixture to at least 100 $^\circ$C. Although the extent of permanent crosslinking was determined primarily by the VTAS concentration, the temperature and duration of the heating were also quite important.

While VTAS is a trifunctional crosslinker, its reactivity decreases as its coordination increases. Its first acetoxy group reacts quickly with a silanol group, even at room temperature, and releases an acetic acid molecule. Its second acetoxy group forms a crosslink more slowly. Its third and final acetoxy group can take hours or days to form a crosslink, even at elevated temperature, and may not form a crosslink at all if the mixture is not heated long enough.

From work on simple borosilicones, it is estimated that 5.8 wt\% VTAS is stoichiometric saturation in 70 cSt STPDMS. Since the gelation threshold for a trifunctional crosslinker is 50\% of saturation, 3 wt\% VTAS should gel 70 cSt STPDMS. In practice, however, VTAS forms its third crosslink so slowly that higher VTAS fractions are needed to approach the gelation threshold in a reasonable amount of time.

The non-simple borosilicone was therefore made by adding 4.5 wt\% VTAS to SFR 70. After blending, the mixture was heated to 130 $^\circ$C for 5 hours to promote crosslinking and evaporation of acetic acid. The resulting partially crosslinked silicone fluid (PCS1) had a room-temperature viscosity of 1.12 Pa$\cdot$s, 16 times that of the original fluid (SFR 70).

PCS1 was vacuum dried to remove residual acetic acid, converted to a borosilicone by adding 1.5 wt\% TMB, and vacuum dried again. It was pressed into a sheet and allowed to equilibrate with laboratory air for 10 days. The resulting non-simple borosilicone, designated NB70-PCS1, is a stiff, transparent liquid with an extremely high viscosity.

\subsection{Stress Relaxation Modulus}

\begin{figure}
	\includegraphics{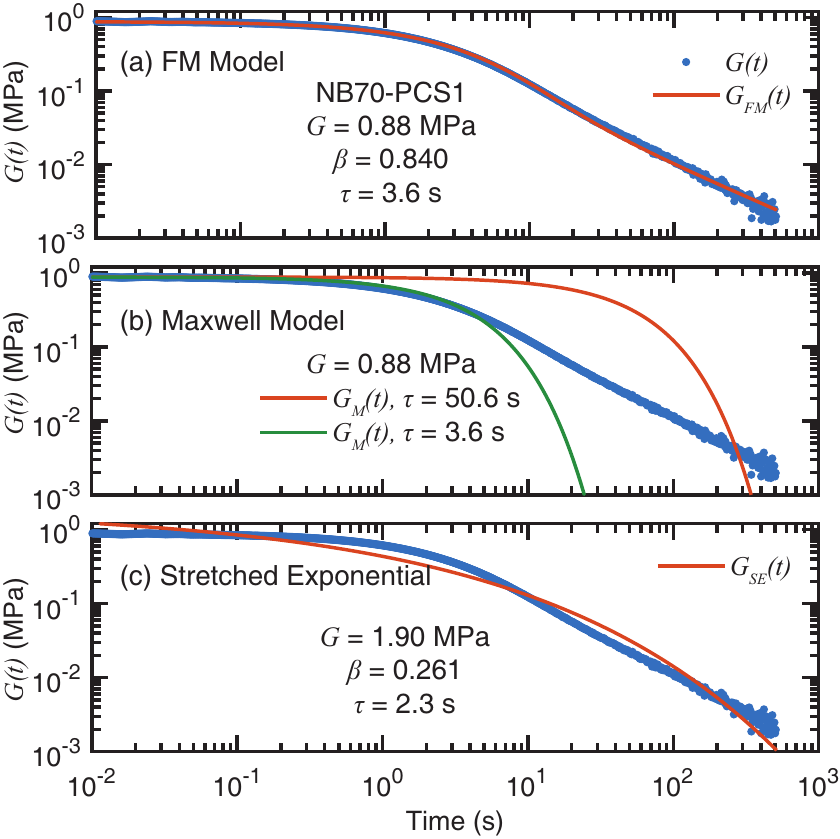}
	\caption{(a) Stress relaxation modulus $G(t)$ of NB70-PCS1, fit by the FM model's stress relaxation modulus $G_{\textsc{fm}}(t)$. Modulus $G$, fractional order $\beta$, and characteristic time $\tau$ are the fit's three parameters. Fits by (b) the Maxwell model's stress relaxation modulus $G_{\textsc{m}}(t)$ and (c) a stretched exponential modulus $G_{\textsc{se}}(t)$ do not match the data well.  \label{fig:NB70-PCS1StressRelaxationModulus}}
\end{figure}

The stress relaxation modulus $G(t)$ of NB70-PCS1 was measured using the sudden compression technique. One such measurement is shown in Fig. \ref{fig:NB70-PCS1StressRelaxationModulus}a, along with a fit by $G_{\textsc{fm}}(t)$, the Fractional Maxwell model's stress relaxation modulus given in Eq. (\ref{eq:FMaxwellStressRelaxation}). The quality of the fit indicates that the Fractional Maxwell model does indeed describe the stress-strain behavior of NB70-PCS1, at least for sudden compression.

The fit provides values for the model's three parameters: the modulus $G$, the fractional order $\beta$, and the characteristic time $\tau$. The modulus $G$ was found to be $0.852\pm0.038$ MPa, somewhat larger than the modulus of SB70-2.60. That result indicates that the presence of permanent crosslinks in the material stiffens its time-average network as compared to a network having purely temporary crosslinks. The characteristic time $\tau$ was found to be 3.6 s, but increased slowly over time as acetic acid from the VTAS gradually dissipated.  

$\beta$ is the fractional order of the FM model's spring-pot and also the random assembly's volume concentration of viscous cuboids. The fit gives $\beta=0.849\pm0.014$, corresponding to a random assembly that is 85\% viscous and 15\% elastic. The effect of $\beta<1$ is to prolong the stress relaxation so that it is slower-than-exponential, as observed in the measurement. For comparison, the exponential relaxation predicted by the Maxwell model does not fit the data (Fig. \ref{fig:NB70-PCS1StressRelaxationModulus}b).

Since the 1854 work of Kohlrausch\cite{kohlrausch1854}, relaxation processes that are slower than exponential have often been fit by stretched exponentials of the form $G_{\textsc{se}}(t)=Ge^{-(t/\tau)^{\beta}}$. Such fits are mostly empirical because few theories predict stretched exponential relaxation\cite{palmer1984}. A connection between theories and the non-simple borosilicones could not be found, which is just as well because $G_{\textsc{se}}(t)$ does not fit the NB70-PCS1 data (Fig. \ref{fig:NB70-PCS1StressRelaxationModulus}c).

\subsection{Shear Viscosity}

The time-dependent shear viscosity $\eta(t)$ of NB70-PCS1 was measured over a range of shear rates using the linear shear technique. Because the results differ qualitatively from those observed in simple borosilicones, they cannot be explained by the transient network and Maxwell models. Instead, they required the FM model and an analysis of how materials exemplifying that model respond to constant shear strain.

As before, the analysis assumes that surfaces perpendicular to axis 2 are sheared along axis 1, so shear stress $\sigma_{\textsc{s}}(t)=\sigma_{21}(t)$. The calculation could proceed using the appropriate constitutive equation, Eq. (\ref{eq:constitutiveFM}), but a faster route is to substitute $G_{\textsc{fm}}(t)$ into Eq. (\ref{eq:genericTDShearViscosity}),
\begin{eqnarray}
\eta_{\textsc{fm}}(t) & = & \int_{0}^{t}GE_{\beta}\left(-\left(\frac{t'}{\tau}\right)^{\beta}\right)dt'\nonumber\\
             & = & GtE_{\beta,2}\left(-\left(\frac{t}{\tau}\right)^{\beta}\right),\label{eq:TDFMShearViscosity}
\end{eqnarray}
where Ref. \cite{mathai2005}, Eq. (2.3.17) has been used to do the integral.

When $\beta = 1$, $\eta_{\textsc{fm}}(t)$ reduces to $\eta_{\textsc{lef}}(t)$ given in Eq. (\ref{eq:LEFTDShearViscosity}). With its only time-dependent term a decaying exponential, $\eta_{\textsc{lef}}(t)$ approaches $\eta_{\textsc{lef}}=G\tau$ at long times.

For $1 > \beta > 0$, however, $\eta_{\textsc{fm}}(t)$ does not approach a constant value at long times. Instead, it behaves as
\begin{equation}
\eta_{\textsc{fm}}(t) = Gt^{1-\beta}\frac{\tau^{\beta}}{\Gamma(2-\beta)}+\mathcal{O}(t^{1-2\beta}),
\end{equation}
where Ref. \cite{podlubny1998}, Eq. (1.143) has been used. The time-dependent viscosity $\eta_{\textsc{fm}}(t)$ of a Fractional Maxwell material thus increases as $Ct^{1-\beta}$ forever.

\begin{figure}
	\includegraphics{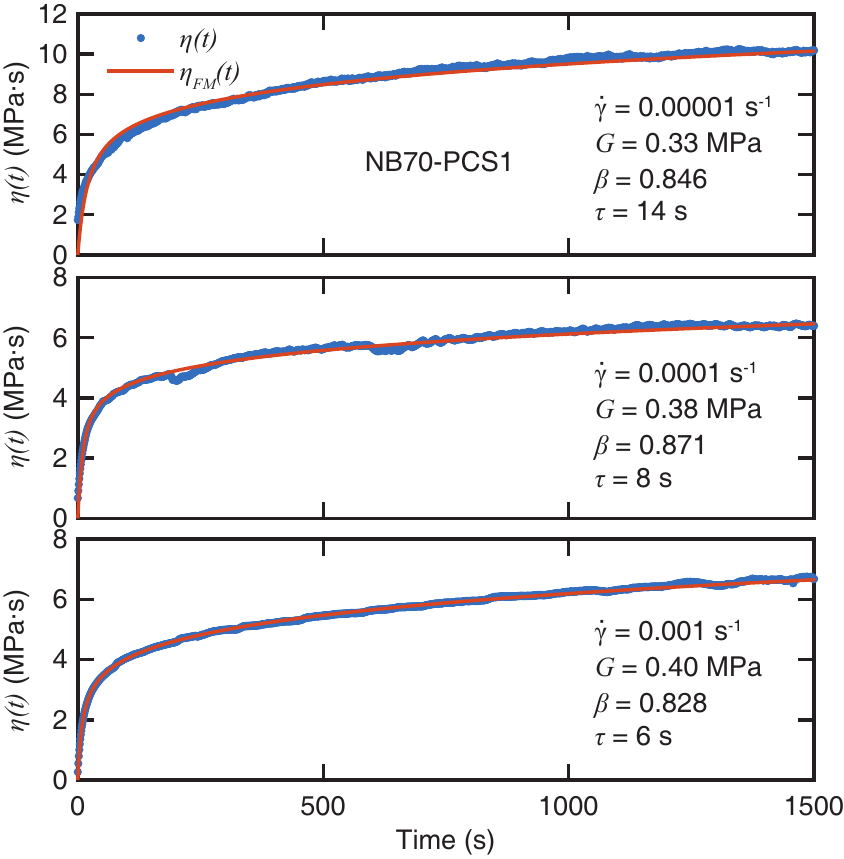}
	\caption{Shear viscosity $\eta(t)$ of NB70-PCS1 at three different shear rates $\dot{\gamma}$. Each measurement is fit by the FM model's shear viscosity $\eta_{\textsc{fm}}(t)$. Modulus $G$, fractional order $\beta$, and characteristic time $\tau$ are the fit's three parameters.\label{fig:figNB70-PCS1ShearViscosity}}
\end{figure}

Measurements of NB70-PCS1's shear viscosity at three different constant shear rates $\dot{\gamma}$ are shown in Fig. \ref{fig:figNB70-PCS1ShearViscosity}, along with fits by $\eta_{\textsc{fm}}(t)$. Those fits yield values for the modulus $G$, the fractional order $\beta$, and the characteristic time $\tau$. Together, the measurements give $\beta = 0.848\pm0.035$, similar to the $\beta$ obtained by the sudden compression technique. As predicted by the FM model, the measured shear viscosities $\eta(t)$ increased with $t$ for as long as the measurements could proceed.

\subsection{Complex Shear Modulus}

\begin{figure}
	\includegraphics{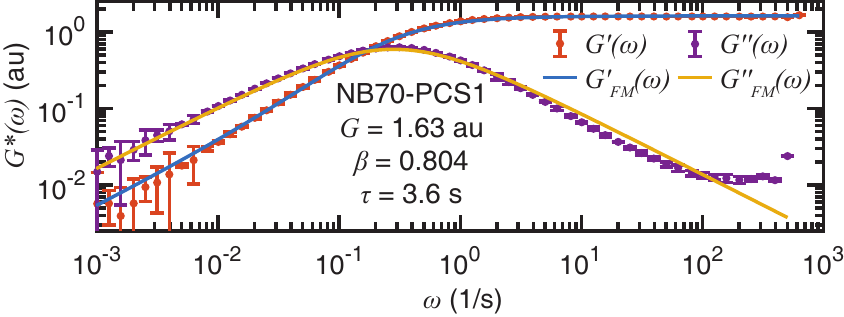}
	\caption{The complex shear modulus $G^{*}(\omega)= G'(\omega) + i G''(\omega)$ of NB70-PCS1, fit by the FM model's complex shear modulus $G_{\textsc{fm}}^{*}(\omega)$.\label{fig:NB70-PCS1ComplexShearModulus}}
\end{figure}

A measurement of the complex shear modulus for NB70-PCS1 is shown in Figure \ref{fig:NB70-PCS1ComplexShearModulus}, along with a fit by $G_{\textsc{fm}}^{*}(\omega)$. The fit is excellent and provides values for the fractional order $\beta=0.804$ and the characteristic time $\tau = 3.6$ s. This $\beta$ is slightly smaller than that obtained via sudden compression or constant shear.

Despite their minor differences, the three measurement techniques agree that NB70-PCS1 is about 15 to 20\% of the way along the path from viscous behavior to elastic behavior. It is definitely a fractional-order viscoelastic material.

\subsection{Other Non-Simple Borosilicones}

\begin{figure}
	\includegraphics{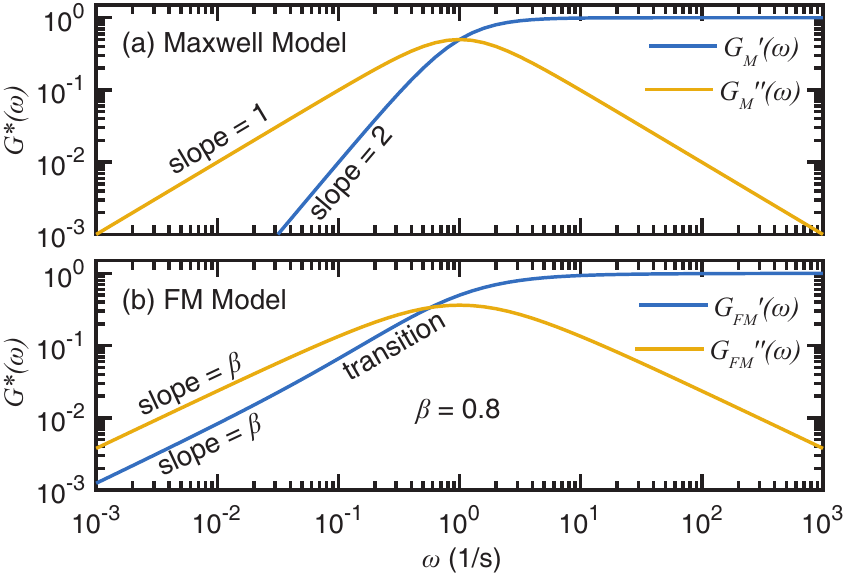}
	\caption{(a) The Maxwell model's complex modulus $G_{\textsc{m}}^{*}(\omega)$ for $G=1$ and $\tau=1$ s. As $\omega\to 0$, $G_{\textsc{m}}'(\omega) \propto \omega^{2}$ and $G_{\textsc{m}}''(\omega) \propto \omega$. (b) The FM model's complex modulus $G_{\textsc{fm}}^{*}(\omega)$ for $G=1$, $\beta=0.8$, and $\tau=1$ s. As $\omega\to 0$, $G_{\textsc{fm}}'(\omega) \propto \omega^{\beta}$ and $G_{\textsc{fm}}''(\omega) \propto \omega^{\beta}$. In the transition region $\cos^{1/\beta}(\pi\beta/2) < \omega\tau < 1$, $G_{\textsc{fm}}'(\omega)$ changes more rapidly than $\omega^{\beta}$.\label{fig:asymptotic}}
\end{figure}

Non-simple borosilicones can be differentiated from simple borosilicones by the behaviors of $G'(\omega)$ and $G''(\omega)$ at small $\omega\tau$. In a simple borosilicone described by the Maxwell model, $G'(\omega) \propto \omega^2$ and $G''(\omega) \propto \omega$ as $\omega\to 0$. In a log-log plot, the slope of $G'(\omega)$ is 2 and the slope of $G''(\omega)$ is 1 (Fig. \ref{fig:asymptotic}).

In a non-simple borosilicone described by the FM model, $G'(\omega) \propto \omega^{\beta}$ and $G''(\omega) \propto \omega^{\beta}$ as $\omega\to 0$. There is also a transition region $\cos^{1/\beta}(\pi\beta/2) < \omega\tau < 1$ in which $G'(\omega)$ changes more rapidly than $\omega^{\beta}$. In a log-log plot, the slopes of $G'(\omega)$ and $G''(\omega)$ are $\beta$ as $\omega\to 0$, but the slope of $G'(\omega)$ exceeds $\beta$ in the transition region. 

Non-simple borosilicones based on partially-crosslinked silicones (e.g., NB70-PCS1) clearly exhibit these reduced asymptotic slopes, but they are not alone. Borosilicones containing reinforcing fillers or silicone gums also exhibit reduced asymptotic slopes and are thus non-simple. 

\begin{figure}
	\includegraphics{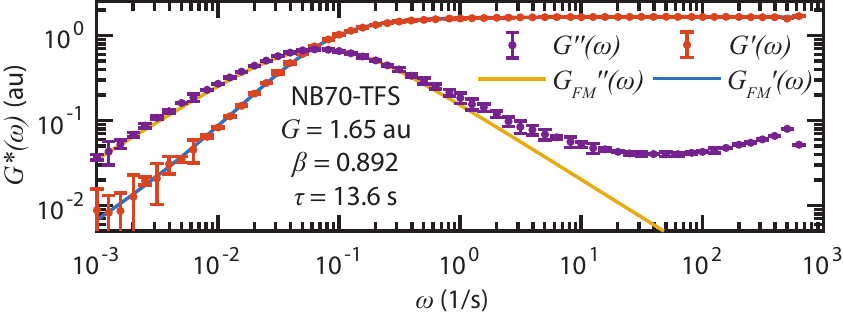}
	\caption{The complex shear modulus $G^{*}(\omega)= G'(\omega) + i G''(\omega)$ of NB70-TFS, fit by the FM model's complex shear modulus $G_{\textsc{fm}}^{*}(\omega)$.\label{fig:NB70-TFSComplexShearModulus}}
\end{figure}

Fig. \ref{fig:NB70-TFSComplexShearModulus} shows $G^{*}(\omega)$ for NB70-TFS, a borosilicone made by adding 40 wt\% Treated Fumed Silica (Cabot TS-530) to SB70-2.60. The silica was milled into the SFR 70 fluid before adding the TMB and vacuum drying. Consistent with the FM model, $G'(\omega)$ and $G''(\omega)$ exhibit reduced slopes as $\omega\to 0$. Moreover, the FM model provides an excellent fit over the entire measured range of $\omega$, apart from some excess loss at high frequencies.

Fumed silica (FS) and its chemically treated version (TFS) are routinely dispersed in silicone rubbers to enhance tensile strength, tear strength, and toughness. The reinforcing nature of these nanometer-scale amorphous silica particles stems from their physical and chemical interactions with the silicone chains. In borosilicones, however, fumed silica evidently plays another important role. The fit of $G^{*}(\omega)$ by $G^{*}_{\textsc{fm}}(\omega)$ suggests that some fraction of the particle-chain interactions act as non-transient couplings so that the RA model applies and NB70-TFS exhibits FM model behavior with $\beta \approx 0.9$.

\begin{figure}
	\includegraphics{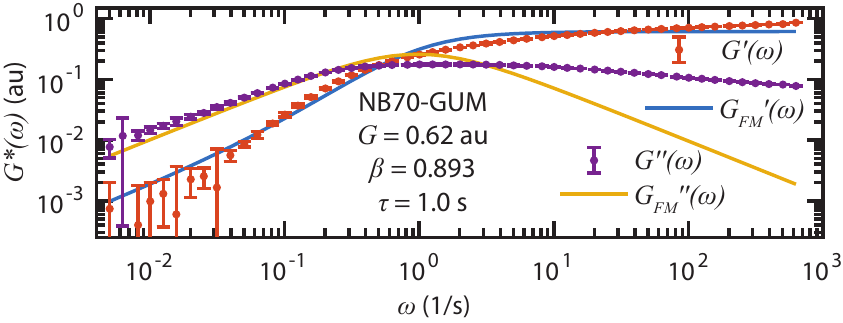}
	\caption{The complex shear modulus $G^{*}(\omega)= G'(\omega) + i G''(\omega)$ of NB70-GUM, fit by the FM model's complex shear modulus $G_{\textsc{fm}}^{*}(\omega)$.\label{fig:NB70-GUMComplexShearModulus}}
\end{figure}

Fig. \ref{fig:NB70-GUMComplexShearModulus} shows $G^{*}(\omega)$ for NB70-GUM, a borosilicone made by adding 50 wt\% 20,000,000 cSt PDMS gum (Gelest DMS-T72) to SB70-2.60. The gum was added to the SFR 70 fluid before adding the TMB and vacuum drying. The slopes of $G'(\omega)$ and $G''(\omega)$ are significantly reduced as $\omega\to 0$, consistent with FM model behavior with $\beta \approx 0.9$. For small $\omega$, $G^{*}(\omega)$ is fit fairly well by $G^{*}_{\textsc{fm}}(\omega)$. For large $\omega$, however, the fit is poor. 

Like fumed silica, silicone gum strengthens silicone rubbers and borosilicones. A gum's long polymer chains disperse stress through their entanglements and extended reaches. When woven into a borosilicone, the gum's chains evidently act as non-transient couplings, so that the RA model applies and results in approximately FM model behavior at small $\omega$. At large $\omega$, however, excess loss first seen in the high-MW borosilicones (Fig. \ref{fig:SBSeriesComplexShearModulus}) obscure much of the FM model behavior.

\subsection{Elongation Viscosity}

\begin{figure}
	\includegraphics{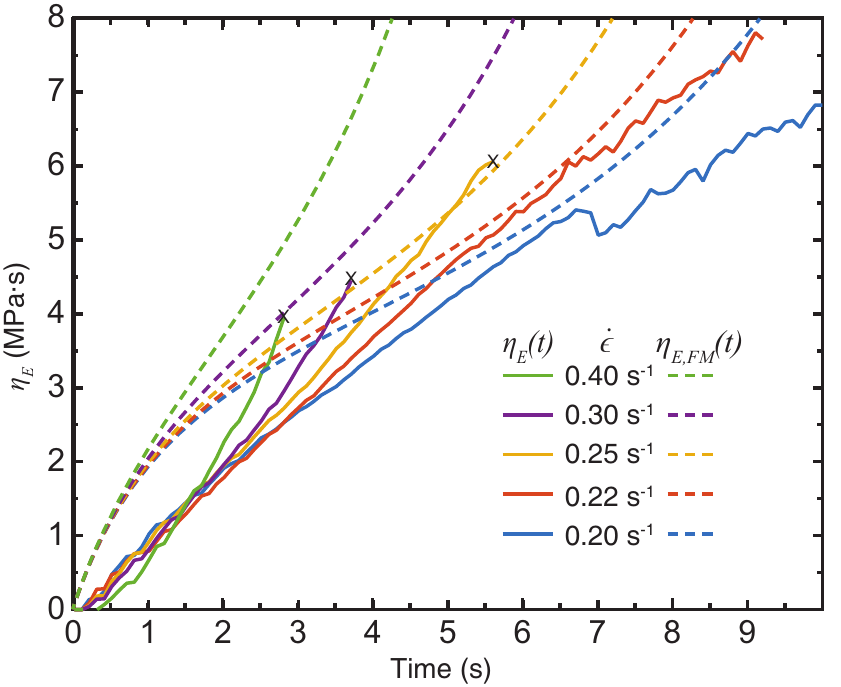}
	\caption{Time-dependent elongation viscosity $\eta_{\textsc{e}}(t)$ curves for NB70-TFS. $\eta_{\textsc{e}}(t)$ was measured at the constant elongation rates $\dot{\epsilon}$ indicated. Breaks of the sample are marked by x. $\eta_{\textsc{e,fm}}(t)$ was calculated numerically from Eq. (\ref{eq:FMElongationViscosity}), using $G=1.05$ MPa, $\beta = 0.893$, and $\tau = 0.796$ s, values obtained by fitting a separate compression measurement of the same NB70-TFS.\label{fig:NB70-TFSElongationViscosity}}
\end{figure}

With its fumed silica reinforcement, NB70-TFS tolerates considerable tensile stress before breaking and its time-dependent elongation viscosity $\eta_{\textsc{e}}(t)$ can be measured at interesting elongation rates. Figure \ref{fig:NB70-TFSElongationViscosity}a shows its $\eta_{\textsc{e}}(t)$ measured at five constant-elongation rates $\dot{\epsilon}$ from 0.20 s${}^{-1}$ to 0.40 s${}^{-1}$, using the exponential stretching technique described earlier.

The tensile strength of NB70-TFS is approximately 1.5 MPa and the beam broke suddenly whenever its tensile stress exceeded that value. Breakage occurred during elongation at the three highest elongation rates. In each break, the beam snapped near its midpoint and left smooth fracture faces, a behavior familiar to anyone who has stretched silicone bouncing putty (e.g., Silly Putty) quickly enough to break it.

The FM model's time-dependent elongation viscosity $\eta_{\textsc{e,fm}}(t)$ was calculated for comparison. For constant-rate elongation $\dot{\epsilon}$, starting at $t = 0$, $\eta_{\textsc{e,fm}}(t)$ can be obtained by substituting $G_{\textsc{fm}}(t)$ into Eq. (\ref{eq:genericTDElongationViscosity}), 
\begin{eqnarray}
\eta_{\textsc{e,fm}}(t) & = & 2G\int_{0}^{t}E_{\beta}(-(t'/\tau)^{\beta})e^{2\dot{\epsilon}t'}dt'\nonumber\\
& & +  G\int_{0}^{t}E_{\beta}(-(t'/\tau)^{\beta})e^{-\dot{\epsilon}t'}dt'.\label{eq:FMElongationViscosity}
\end{eqnarray}

Closed form solutions for these integrals could not be found, but it was possible to determine their behaviors as $t\to\infty$. The second integral converges to a finite value,
\begin{equation}
G\tau\frac{(\dot{\epsilon}\tau)^{\beta-1}}{1+(\dot{\epsilon}\tau)^{\beta}},
\end{equation}
where Ref. \cite{haubold2011}, Eq. (7.1) has been used. The first integral diverges with a leading term approximated using Ref. \cite{podlubny1998}, Eq. (1.143) as
\begin{equation}
\frac{2G\tau^{\beta}}{\Gamma(1-\beta)(1-\beta)} 
t^{1-\beta} {}_{1}F_{1}(1-\beta;2-\beta;2\dot{\epsilon}t),
\end{equation}
where ${}_1F_{1}(a;b;z)$ is the Kummer confluent hypergeometric function. The elongation viscosity $\eta_{\textsc{e,fm}}(t)$ therefore grows without limit as $t\to\infty$ and any borosilicone described by the FM model must eventually break when subject to constant-rate elongation.

For finite $t$, $\eta_{\textsc{e,fm}}(t)$ could only be computed numerically. Unfortunately, fits using the computational $\eta_{\textsc{e,fm}}(t)$ were unsuccessful, so the model's parameters $G=1.05$ MPa, $\beta=0.893$, and $\tau=0.80$ s were found instead using the sudden compression technique on the same portion of NB70-TFS. The $\eta_{\textsc{e,fm}}(t)$ curves shown in Fig \ref{fig:NB70-TFSElongationViscosity} were then calculated numerically at the five constant elongation rates $\dot{\epsilon}$.

The agreement between $\eta_{\textsc{e}}(t)$ and $\eta_{\textsc{e,fm}}(t)$ is far from perfect, with initial slopes differing by about a factor of 2. Nonetheless, measurement and model exhibit significant similarities in scale and curvature, particularly given that the $\eta_{\textsc{e,fm}}(t)$ curves involved no parameter adjustments. 

\subsection{Revisiting Simple Borosilicones}

\begin{figure}
	\includegraphics{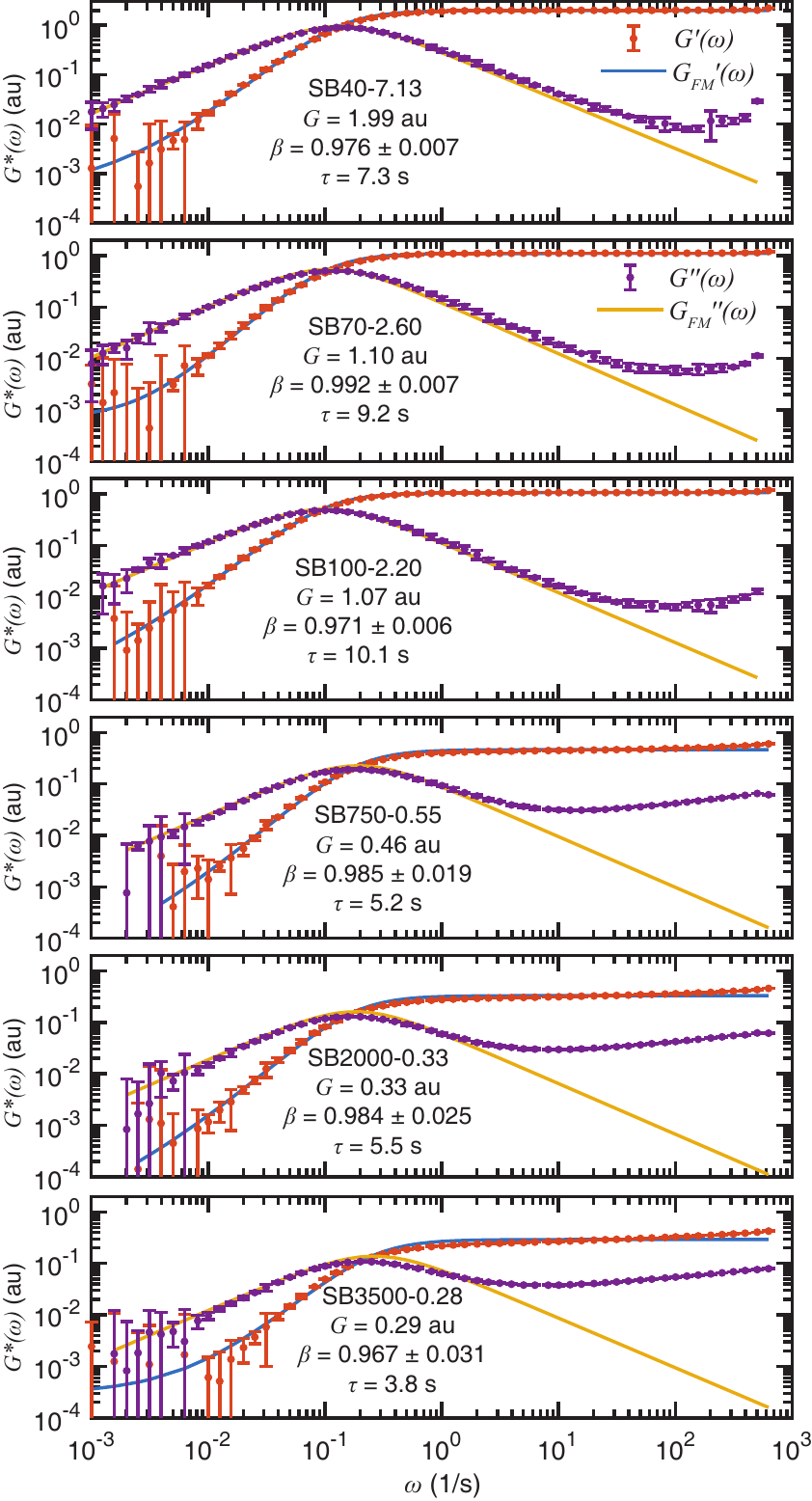}
	\caption{Complex shear moduli $G^{*}(\omega)= G'(\omega) + i G''(\omega)$ of simple borosilicones SB70-2.60 through SB3500-0.28, each fit by the FM model's complex shear modulus $G^{*}_{\textsc{fm}}(\omega)$. The fit's 95\% confidence interval is given for $\beta$.\label{fig:SBSeriesFMComplexShearModulus}}
\end{figure}

While simple borosilicones exemplify the Lodge Elastic Fluid concept and their measured complex shear moduli are well fit by the Maxwell model's complex shear modulus $G_{\textsc{m}}^{*}(\omega)$ (Fig. \ref{fig:SBSeriesComplexShearModulus}), the same data can be fit even more tightly using the Fractional Maxwell model's complex shear modulus $G_{\textsc{fm}}^{*}(\omega)$ (Fig. \ref{fig:SBSeriesFMComplexShearModulus}). Of course, adding another parameter to a fit can improve that fit even when the added parameter has no physical justification. In this case, however, including the fractional order $\beta$ is a recognition that simple borosilicones are not perfect systems and could have some non-transient couplings between their polymer chains.

If there were no non-transient couplings in the simple borosilicones, the fits to data should find $\beta=1$ and the FM model should reduce to the Maxwell model. That outcome is not observed. Instead, the fits shown in Fig. \ref{fig:SBSeriesFMComplexShearModulus} consistently yield $\beta$ just less than 1 and, in four of the six fits, the 95\% confidence interval excludes $\beta=1$.  There appear to be small but non-zero concentrations of non-transient couplings in simple borosilicones, due perhaps to steric hindrances, impurities, or clustering of boron and oxygen atoms. 

\subsection{The Approach to Solid}

The RA and FM models both predict liquid behavior, $G^{*}(0)=0$ and $G(\infty)=0$, when $\beta > 0$. One might therefore expect that, as the fraction of permanent crosslinks in a non-simple borosilicone increases almost to one, the borosilicone will remain liquid and its $\beta$ will decrease almost to zero. 

In reality, however, when $\beta$ drops below about 0.8, something not predicted by the RA or FM models occurs: the borosilicone begins to exhibit solid behavior, $G^{*}(0)\ne 0$ and $G(\infty)\ne 0$. The RA and FM models, which do not consider real crosslinks between real polymer chains, fail to predict the polymer phase transition known as gelation.

A simple borosilicone is a network liquid composed of silicone chains connected only by temporary crosslinks. When a few permanent crosslinks are added, they create some permanently-crosslinked subnetworks---that is, parts of the overall network that cannot come apart.

As long as the permanent crosslink concentration is low, those subnetworks are small and isolated, and connect to one another only through temporary crosslinks. The borosilicone is a sol (or solution) of permanently-crosslinked subnetworks in an otherwise temporarily-crosslinked liquid borosilicone.

Once the concentration of permanent crosslinks exceeds the gelation threshold, however, the permanently-crosslinked subnetworks are no longer isolated. Instead, many or most of them coalesce into an enormous permanently-crosslinked network that spans the material and prevents it from fully relaxing stress, even as $t\to\infty$.

The borosilicone is then a gel and a network solid. Its permanent crosslinks give a fixed aspect to its overall network topology and its temporary crosslinks add a temporary aspect as well. In effect, this type of borosilicone is a network liquid piggybacking on the framework of a network solid. 

To find the fractional order $\beta$ associated with the gelation threshold, a liquid borosilicone was produced as close as possible to that threshold. Designated NB70-PCS2, it was based on PCS2, a nearly-gelled partially-crosslinked silicone fluid.

PCS2 was made by adding 4.0 wt\% VTAS to SFR 70 and heating the mixture to 150 ${}^{\circ}$C until it had almost gelled. When, after more than 30 hours, the liquid had become so viscous that gelation was at most minutes away, it was cooled to room temperature to stop the crosslinking reaction.

The viscosity of this PCS2 was 26.3 Pa$\cdot$s, approximately 400 times the viscosity of the original fluid (SFR 70). PCS2 was then converted into a borosilicone by adding 1.50 wt\% TMB and vacuum drying. The resulting NB70-PCS2 was pressed into a rough sheet and allowed to equilibrate with laboratory air for 10 days.

\begin{figure}
	\includegraphics{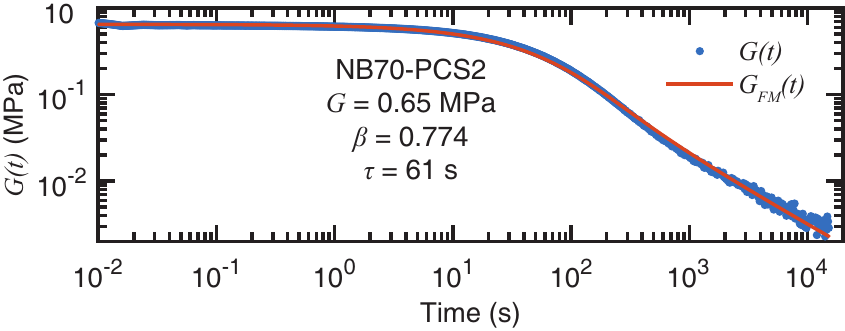}
	\caption{Stress relaxation modulus $G(t)$ of NB70-PCS2, fit by the FM model's stress relaxation modulus $G_{\textsc{fm}}(t)$. \label{fig:NB70-PCS2StressRelaxationModulus}}
\end{figure}

\begin{figure}
	\includegraphics{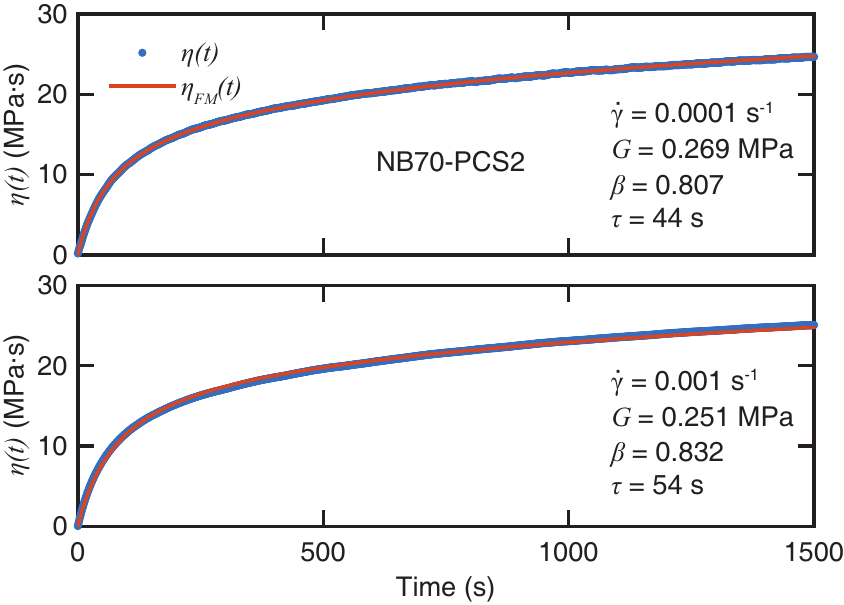}
	\caption{Shear viscosity $\eta(t)$ of NB70-PCS2 at two different shear rates $\dot{\gamma}$. Each measurement is fit by the FM model's shear viscosity $\eta_{\textsc{fm}}(t)$.\label{fig:figNB70-PCS2ShearViscosity}}
\end{figure}

\begin{figure}
	\includegraphics{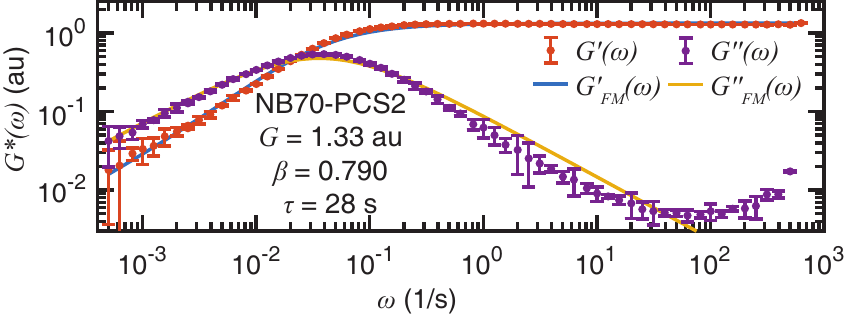}
	\caption{The complex shear modulus $G^{*}(\omega)= G'(\omega) + i G''(\omega)$ of NB70-PCS2, fit by the FM model's complex shear modulus $G_{\textsc{fm}}^{*}(\omega)$.\label{fig:NB70-PCS2ComplexShearModulus}}
\end{figure}

The properties of the dry NB70-PCS2 were then measured and fit to the FM model, with the primary goal of determining the fractional order $\beta$. The stress relaxation modulus $G(t)$ (Fig. \ref{fig:NB70-PCS2StressRelaxationModulus}) was measured repeatedly and fit by $G_{\textsc{fm}}(t)$ to obtain $\beta = 0.785\pm0.054$. The shear viscosity $\eta(t)$ (Fig. \ref{fig:figNB70-PCS2ShearViscosity}) was measured at two constant shear rates $\dot{\gamma}$ and fit by $\eta_{\textsc{fm}}(t)$ to obtain $\beta=0.820\pm0.025$. And the complex shear modulus $G^{*}(\omega)$ (Fig. \ref{fig:NB70-PCS2ComplexShearModulus}) was measured and fit by $G_{\textsc{fm}}^{*}(\omega)$ to obtain $\beta=0.790$. 

Based on these values, $\beta$ at the gelation threshold is estimated as $0.80\pm0.03$. That threshold value, however, probably depends somewhat on the detailed network structure of borosilicone, particularly the functionality and coordination of the crosslinkers. Moreover, when the three approaches to making non-simple borosilicones (partial crosslinking, adding TFS, and adding silicone gum) are combined in a single non-simple borosilicone, the observed $\beta$ value is somewhat smaller than if one of those approaches had been omitted.

What is clear, however, is that preparing liquid borosilicones with $\beta < 0.8$ is difficult. Smaller $\beta$ values require greater permanent-crosslink concentrations and can easily lead to gelation. A more promising path to small $\beta$ values is to stop trying to avoid gelation. Instead, cross the gelation threshold deliberately and thereby produce solid borosilicones. 

\section{Viscoelastic Silicone Rubber} 

Preparing partially-crosslinked silicone fluids close to the gelation threshold is difficult and most attempts resulted in sticky, gelatinous goos. PCS2 was made in the attempt that came closest to the threshold without actually it.

Once a partially-crosslinked silicone has crossed the threshold and gelled, turning it into a homogeneous borosilicone is nearly impossible. To produce borosilicones beyond the gelation threshold, the TMB must be added while the partial-crosslinking is still in progress and before gelation has occurred. Even that approach has difficulties because TMB interferes with the partial-crosslinking process and stiffens the still-crosslinking material.

Despite those complications, partial-crosslinking can produce borosilicones beyond the gelation threshold\cite{bloomfield2015}. With their material-spanning permanent networks, these solid borosilicones cannot flow in response to stress and will be referred to as borosilicone rubbers or viscoelastic silicone rubbers (VSRs).

To produce a VSR that is free of behavior-influencing catalysts, the partial crosslinking was done using VTAS and high temperatures. The starting fluid for this preparation was PCS1, the partially-crosslinked fluid used to make NB70-PCS1. Since PCS1 is already close to the gelation threshold, it can be driven across that threshold by a small increase in the concentration of permanent crosslinks.

To 15g of PCS1 were added 4.5 wt\% VTAS and 1.11 wt\% TMB, followed immediately by 5 seconds of vigorous stirring and 30 seconds of vacuum drying. The mixture contained enough permanent crosslinker (VTAS) to well-exceed the gelation threshold. It also contained enough temporary crosslinker (TMB) to become a borosilicone. It began thickening immediately.

Before the mixture could gel, it was transferred into two aluminum disk molds (38 mm dia x 6.35 mm) and heated under pressure to 200 $^{\circ}$C for five minutes. The high temperature greatly increased the permanent crosslinking rate, allowing the material to solidify into disks of viscoelastic silicone rubber, designated VSR70-PCS.

To prevent bubble formation, the VSR70-PCS disks were cooled under pressure to room temperature. Once removed from their molds, the disks were transparent and bubble-free, but they contained methanol and acetic acid derived from the VTAS and TMB. To eliminate those volatile compounds and their esters, the disks were dried in 70 $^{\circ}$C air for several weeks. 

\subsection{The Fractional Zener Model}

VSR70-PCS is a solid with viscoelastic properties like a non-simple borosilicone. To describe its behavior, a viscoelastic model must combine the fractional-order stress-strain response of a non-simple borosilicone for $\omega > 0$ and $t < \infty$ and the stress-strain response of a solid at $\omega=0$ and $t \to \infty$. The simplest viscoelastic model that satisfies those requirements is the Fractional Zener model.

\begin{figure}
	\includegraphics{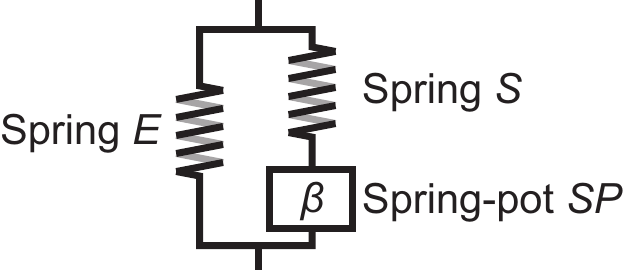}%
	\caption{The Fractional Zener Model consists of a spring $E$ in parallel with the Fractional Maxwell model, which itself consists of spring $S$ in series with spring-pot $SP$. \label{fig:FractionalZenerModel}}
\end{figure}

Shown in Fig. \ref{fig:FractionalZenerModel}, the Fractional Zener model consists of two components in parallel: spring $E$ and a Fractional Maxwell component. The Fractional Maxwell component itself consists of two components in series: spring $S$ and spring-pot $SP$.

The properties of the Fractional Zener (FZ) model can be calculated from those of its two component. Spring $E$ has $G_{\textsc{e}}^{*}(\omega)=E$ and $G_{\textsc{e}}(t)=E$. The FM component has $G^{*}_{\textsc{fm}}(\omega)$ given in Eq. (\ref{eq:FractionalMaxwell}) and $G_{\textsc{fm}}(t)$ given in Eq. (\ref{eq:FMaxwellStressRelaxation}). Using the parallel formula, the FZ model's complex modulus $G^{*}_{\textsc{fz}}(\omega)$ is
\begin{eqnarray}
G^{*}_{\textsc{fz}}(\omega) & = & G^{*}_{\textsc{e}}(\omega) + G^{*}_{\textsc{fm}}(\omega)\nonumber\\
& = & E + G\frac{(i\omega\tau)^{\beta}}{1+(i\omega\tau)^{\beta}}\label{eq:FZComplexModulus}
\end{eqnarray}
and its stress relaxation modulus
\begin{eqnarray}
G_{\textsc{fz}}(t) & = & G_{\textsc{e}}(t) + G_{\textsc{fm}}(t)\nonumber\\
& = & E + GE_{\beta}\left(-(t/\tau)^{\beta}\right)\label{eq:FZStressRelaxationModulus}
\end{eqnarray}

It will be shown that the FZ model's predictions fit the measured behaviors of viscoelastic silicone rubbers remarkably well. While there are other fractional-order viscoelastic models for solids, they are not needed here.

\subsection{Stress Relaxation Modulus}

\begin{figure}
	\includegraphics{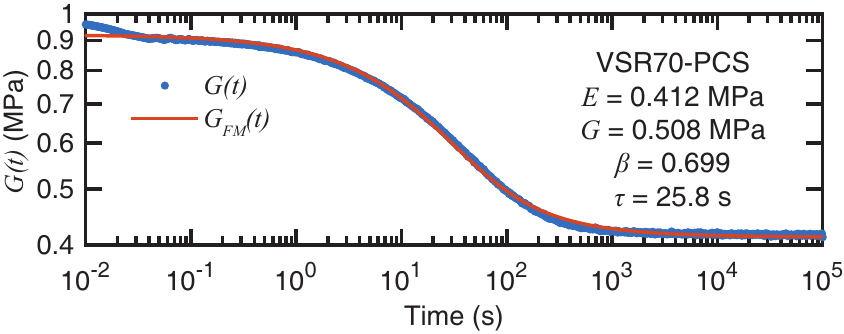}
	\caption{Stress relaxation modulus $G(t)$ of VSR70-PCS, fit by the FZ model's stress relaxation modulus $G_{\textsc{zm}}(t)$. Elastic modulus $E$, modulus $G$, fractional order $\beta$, and characteristic time $\tau$ are the fit's four parameters.\label{fig:VSR70-PCSStressRelaxationModulus}}
\end{figure}

The stress relaxation modulus $G(t)$ for VSR70-PCS was measured using the sudden compression technique. One such measurement is shown in Fig. \ref{fig:VSR70-PCSStressRelaxationModulus}, along with a fit by $G_{\textsc{fz}}$. The fit is excellent over 7 orders of magnitude in time, demonstrating how well the FZ model describes the stress-strain behavior of VSR70-PCS under compression.

The fit to each measurement of $G(t)$ provided values for the model's four parameters: the elastic modulus $E$ associated with the spring component, the FM modulus $G$ associated with the FM component, the fractional order $\beta$, and the characteristic time $\tau$.

VSR70-PCS's elastic modulus is $E=0.391\pm 0.102$ MPa. $E$ represents the stiffness of VSR70-PCS's permanent network, present at all timescales and visible in Fig. \ref{fig:VSR70-PCSStressRelaxationModulus} as VSR70-PCS's modulus at large times.

VSR70-PCS's FM modulus is $G=0.539\pm 0.072$ MPa. $G$ reflects the added stiffness present immediately after compression due to the temporary network piggybacking on the permanent network, the network liquid piggybacking on the network solid. It is visible in Fig. \ref{fig:VSR70-PCSStressRelaxationModulus} as the difference between VSR70-PCS's modulus at small times and its modulus at large times.

Immediately after compression, the temporary network contributes its modulus $G$ to the stiffness of VSR70-PCS as though its crosslinks were permanent. The instantaneous modulus $G(0)$ is thus the sum of the individual moduli $G(0)=E+G$. With increasing time, however, the temporary network relaxes and VSR70-PCS's stiffness decreases to $G(\infty) = E$. The instantaneous modulus has a temporary fraction $G/(E+G) = 0.58$ that decays with time and a permanent fraction $E/(E+G) = 0.42$ that remains indefinitely.

VSR70-PCS's fractional order is $\beta=0.679\pm 0.024$, significantly smaller than that observed in the liquid borosilicones. Although the RA model does not predict gelation, it does predict a decreasing $\beta$ as the volume concentration of viscous cuboids decreases and that of elastic cuboids increases. Compared to any of the liquid borosilicones, the concentration of temporary crosslinks in VSR70-PCS is smaller and the concentration of permanent crosslinks is larger, so VSR70-PCS's smaller $\beta$ is consistent with the RA model's predictions. 

\subsection{Complex Shear Modulus}

\begin{figure}
	\includegraphics{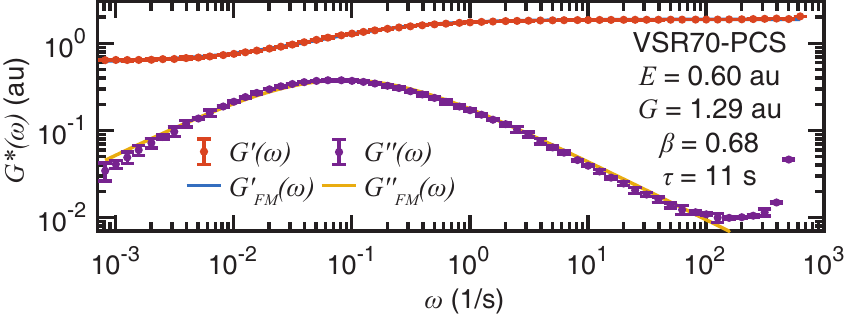}
	\caption{The complex shear modulus $G^{*}(\omega)= G'(\omega) + i G''(\omega)$ of VSR70-PCS, fit by the FZ model's complex shear modulus $G_{\textsc{fm}}^{*}(\omega)$.\label{fig:VSR70-PCSComplexShearModulus}}
\end{figure}

A measurement of the complex shear modulus for VSR70-PCS is shown in Fig. \ref{fig:VSR70-PCSComplexShearModulus}, along with a fit to $G^{*}_{\textsc{fz}}(\omega)$. The fit gives the fractional order $\beta=0.678\pm0.017$, consistent with the value obtained from the $G(t)$ measurement.

Although the elastic modulus $E$ and FM modulus $G$ are reported in arbitrary units, the instantaneous modulus has temporary fraction $G/(E+G)=0.68$ and permanent fraction $E/(E+G)=0.32$. Those two values differs somewhat from the values obtained from $G(t)$. The origin of that difference is not yet understood.

\subsection{Other Viscoelastic Silicone Rubbers}

To prepare VSRs with even higher permanent-crosslink concentrations, a second crosslinking technique was used. In addition to forming permanent crosslinks, vinyltriacetoxylsilane incorporates vinyl groups (Si-CH=CH$_2$) into the non-simple borosilicone's subnetworks. When a silicone fluid containing hydride groups (Si-H) is added to that borosilicone, hydride groups can undergo addition reactions with vinyl groups, forming permanent crosslinks (Si-CH$_2$-CH$_2$-Si) between the silicone chains. With the help of these hydride-vinyl crosslinks, it is possible to exceed the gelation threshold and transform the liquid borosilicones in a viscoelastic silicone rubber\cite{bloomfield2017}. 

The addition reaction is easily controlled with catalyst and temperature, making it possible to prepare and dry a liquid borosilicone containing vinyl and hydride group and then cure that liquid into a viscoelastic silicone rubber. With a platinum catalyst and elevated temperatures, the curing process can take minutes or less. Although the platinum catalyst remains in the finished VSR, it is present in such minuscule concentration that it has no observable effects on the resulting VSR.

Since PCS1 was partially crosslinked using VTAS, it contains a high concentration of vinyl groups and can be cured into VSR via the addition reaction. Three different VSRs (Table {\ref{tab:VSRs}}) were prepared by combining PCS1 with TMB, a hydride-containing polysiloxane, and a platinum catalyst, followed by drying and heating to 200 $^\circ$C for 5 minutes. The hydride-containing polysiloxane was Masil XL-1 (Dystar), a copolymer with approximately 25\% hydromethylsiloxane monomers. The platinum catalyst was Wacker Batch Aux 2 Pt.

\begin{table}
	\caption{Compositions of four VSRs produced using the addition reaction. 0.225g Wacker Batch Aux 2 Pt catalyst was added to each composition before vacuum drying and curing at 200 $^{\circ}$C for 5 minutes.\label{tab:VSRs}}
	\begin{ruledtabular}
		\begin{tabular}{lcccc}
			\textrm{Name}&
			\textrm{PCS1 (g)}&
			\textrm{TMB (g)}&
			\textrm{SF 201 (g)}&
			\textrm{XL-1 (g)}\\
			\colrule
			VSR70-A1 & 15.000 & 0.225 & 0.000 & 0.375\\
			VSR70-A2 & 11.250 & 0.169 & 3.750 & 0.337\\
			VSR70-A3 & 7.500 & 0.112 & 7.500 & 0.300\\
		\end{tabular}
	\end{ruledtabular}
\end{table}

In two of the VSRs, some of the borosilicone was replaced by 1000 cSt vinyl-terminated PDMS (Masil SF 201, Dystar). This substitution diluted the borosilicone and lowered the concentration of temporary crosslinks. VSR70-A1 is undiluted while VSR70-A2 is 75\% borosilicone and VSR70-A3 is 50\% borosilicone.

\begin{figure}
	\includegraphics{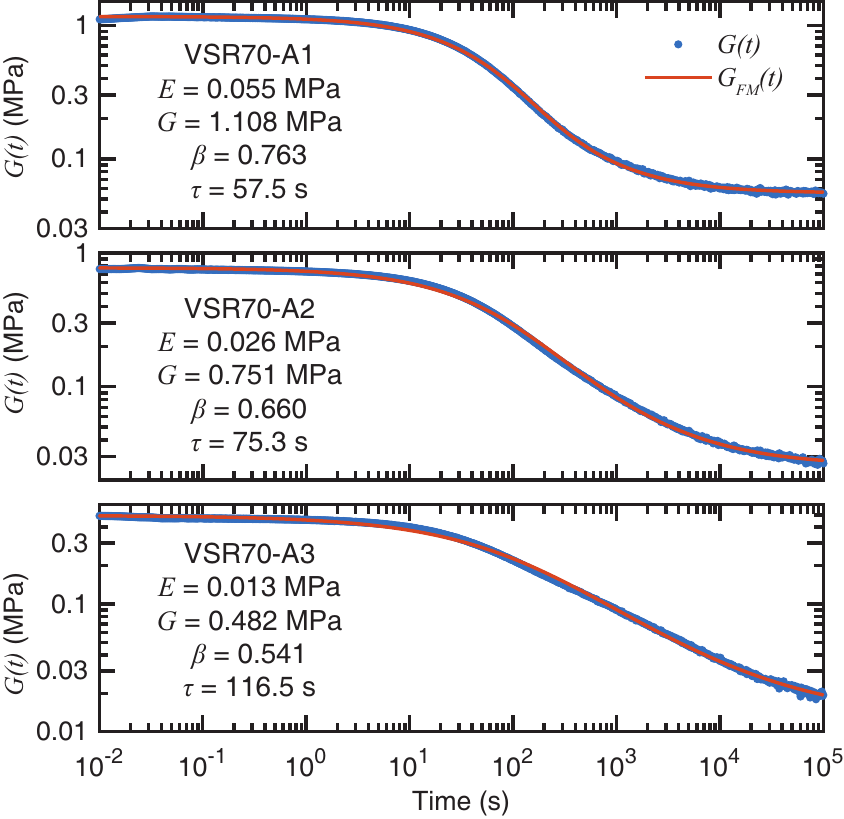}
	\caption{Stress relaxation modulus $G(t)$ of VSR70-A1 through A3, fit by the FZ model's stress relaxation modulus $G_{\textsc{fz}}(t)$. Elastic modulus $E$, modulus $G$, fractional order $\beta$, and characteristic time $\tau$ are the fit's four parameters.\label{fig:VSR70-ASeriesStressRelaxationModulus}}
\end{figure}

Figure \ref{fig:VSR70-ASeriesStressRelaxationModulus} shows the stress relaxation modulus $G(t)$ measured for VSR70-A1 through VSR70-A3, along with fits by $G_{\textsc{fz}}(t)$. Those fits yield values for the elastic modulus $E$, FM modulus $G$, fractional order $\beta$, and characteristic time $\tau$. Many such measurements were made and the values obtained are summarized in Table \ref{tab:VSRsStressRelaxationModulus}. $\tau$, which depends sensitively on chemical environment, is omitted.

\begin{table}
	\caption{Elastic modulus $E$, FM modulus $G$, and fractional order $\beta$ obtained by fitting the FZ model's stress relaxation modulus $G_{\textsc{fz}}(t)$ to stress relaxation modulus $G(t)$ for multiple measurements of VSR70-A1 through VSR70-A3.\label{tab:VSRsStressRelaxationModulus}}
	\begin{ruledtabular}
		\begin{tabular}{lccc}
			\textrm{Name}&
			E\textrm{ (MPa)}&
			G\textrm{ (MPa)}&
			$\beta$\\
			\colrule
			VSR70-A1 & $0.061\pm0.009$ & $1.065\pm0.200$ & $0.777\pm0.031$\\
			VSR70-A2 & $0.030\pm0.013$ & $0.721\pm0.116$ & $0.677\pm0.067$\\
			VSR70-A3 & $0.024\pm0.023$ & $0.397\pm0.092$ & $0.570\pm0.132$\\
		\end{tabular}
	\end{ruledtabular}
\end{table}

As the borosilicone fraction decreases from 100\% in VSR70-A1, to 75\% in VSR70-A2, to 50\% in VSR70-A3, the concentration of temporary crosslinks decreases and so do the fractional order $\beta$, the FM modulus $G$, and the elastic modulus $E$. The decrease in $\beta$ is consistent with the RA model, which predicts such a decrease as the fractional concentration of viscous cuboids decreases.

$G$ is the modulus of the temporary network, the network liquid piggybacking on the network solid. That $G$ decreases almost in proportion to VSR's borosilicone fraction indicates that $G$ is due primarily to that borosilicone fraction. That is expected since, without the borosilicone and its temporary crosslinks, there would be no Fractional Maxwell behavior.

$E$ is the modulus of the permanent network, the network solid. In these VSRs, $E$ decreases as the borosilicone fraction decreases because SF 201 forms a softer permanent network that does PCS1. The average molecule in PCS1 has multiple vinyl groups linked by short PDMS chains, whereas an SF 201 molecule has two vinyl groups linked by a long PDMS chain. Reducing the borosilicone fraction thus has the side effect of weakening the permanent network and decreasing $E$.

\begin{figure}
	\includegraphics{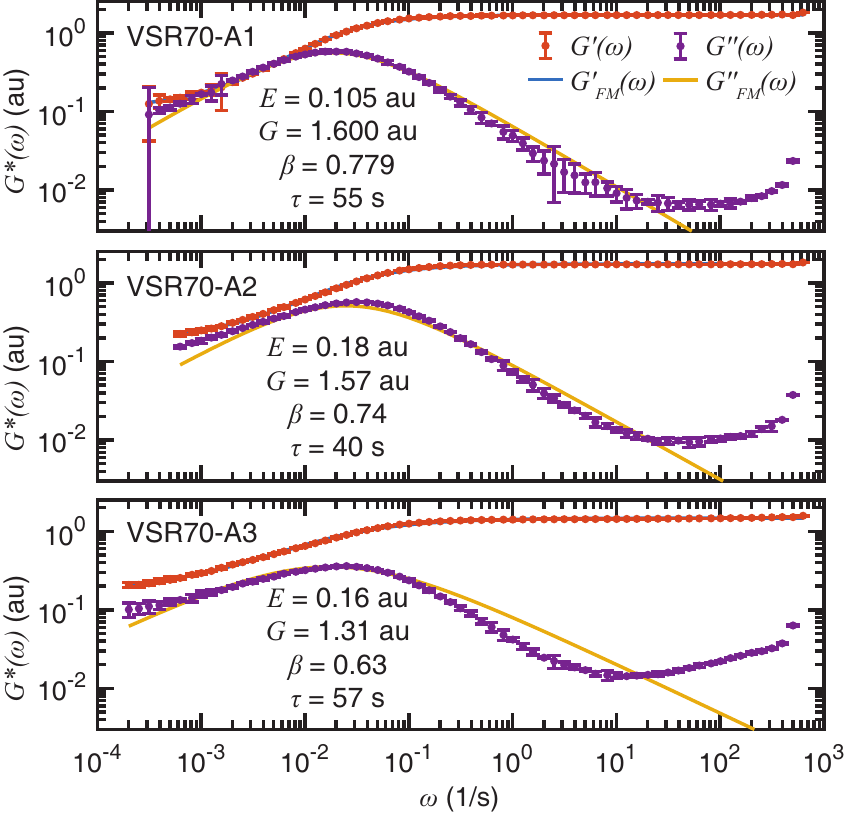}
	\caption{The complex shear modulus $G^{*}(\omega)= G'(\omega) + i G''(\omega)$ of VSR70-A1 through A3, fit by the FZ model's complex shear modulus $G_{\textsc{fz}}^{*}(\omega)$. Elastic modulus $E$, modulus $G$, fractional order $\beta$, and characteristic time $\tau$ are the fit's four parameters.\label{fig:VSR70-ASeriesComplexShearModulus}}
\end{figure}

Figure \ref{fig:VSR70-ASeriesComplexShearModulus} shows the complex shear modulus $G^{*}(\omega)$ for those same three VSRs, along with fits by $G_{\textsc{fz}}^{*}(\omega)$. Elastic modulus fraction $E/(E+G)$, FM modulus fraction $G/(E+G)$, and fractional-order $\beta$ obtained from multiple measurements are shown in Table \ref{tab:VSRsComplexShearModulus}.

\begin{table}
	\caption{Elastic modulus fraction $E/(E+G)$, FM modulus fraction $G/(E+G)$, and fractional-order $\beta$ obtained by fitting the FZ model's complex modulus $G^{*}_{\textsc{fz}}(\omega)$ to complex shear modulus $G^{*}(\omega)$ for multiple measurements of VSR70-A1 through VSR70-A3.\label{tab:VSRsComplexShearModulus}}
	\begin{ruledtabular}
		\begin{tabular}{lccc}
			\textrm{Name}&
			E/(E+G)&
			G/(E+G)&
			$\beta$\\
			\colrule
			VSR70-A1 & $0.062\pm0.003$ & $0.938\pm0.003$ & $0.777\pm0.007$\\
			VSR70-A2 & $0.086\pm0.005$ & $0.914\pm0.005$ & $0.732\pm0.012$\\
			VSR70-A3 & $0.132\pm0.008$ & $0.868\pm0.008$ & $0.631\pm0.015$\\
		\end{tabular}
	\end{ruledtabular}
\end{table}

For VSR70-A1 (100\% borosilicone), the fractional order $\beta$ and the modulus fractions $E/(E+G)$ and $G/(E+G)$ agree well with values obtained from the stress relaxation modulus $G(t)$.

For VSR70-A2 (75\% borosilicone) and VSR70-A3 (50\% borosilicone), $\beta$ agrees with the value obtained from the stress relaxation modulus $G(t)$, but modulus fractions do not. In both cases, the complex shear modulus measurements observe a somewhat larger elastic modulus fraction than is observed in the stress relaxation measurements. That disagreement is probably due in part to the presence of two very different PDMS chain lengths in these diluted-borosilicone VSRs.

Beyond basic science, viscoelastic silicone rubbers exhibit shape memory behaviors that make them useful in practical applications. Because a VSR's network liquid component opposes sudden changes in shape, it takes time for the VSR to ``learn'' a new shape or ``forget'' a previous shape. When subject to external-imposed change in strain, the VSR responds with a large initial stress that gradually relaxes as the VSR complies with the new strain. When left alone, the VSR's network solid component gradually returns the VSR to the shape in which it was originally formed.

\section{About the Temporary Crosslinks}

Each model used to describe a borosilicone or VSR has a characteristic time $\tau$. The accuracy with which the models describe the behaviors of the materials suggests that $\tau$ is closely related to some physical process in the materials themselves.

In the TN model, from which the other models eventually emerged, $\tau$ is the mean lifetime of a strand. Given that temporary crosslinks emulate strand-breaking, it is reasonable to suppose that $\tau$ arises from and is proportional to the mean lifetime of a temporary crosslink $\tau_{\textsc{tc}}$,
\begin{equation}
\tau = a_{\textsc{b}}\tau_{\textsc{tc}},\label{eq:taus}
\end{equation}
where $a_{\textsc{b}}$ may depend on the network structure of the borosilicone or VSR. Given that simple borosilicones have only temporary network structure, it is likely that $a_{\textsc{b}} \approx 1$. Because of their permanent subnetworks and networks, non-simple borosilicones and VSRs have more difficulty relaxing stress and probably have $a_{\textsc{b}} > 1$.

When a simple borosilicone has been cleanly prepared and vacuum-dried, $\tau$ can be 40 s or more. When that same borosilicone is allowed to equilibrate with laboratory air (20 $^\circ$C, 50\% relative humidity), $\tau$ is typically 15 s or less. Since the water concentration in PDMS rubber equilibrated with 20 $^\circ$C, 50 \%RH air is only $\sim$110 ppm ($\sim$0.011 wt\%),\cite{barrie1969} water evidently reduces the mean lifetime of the temporary crosslinks catalytically.

The catalysis mechanism probably involves exchange reactions, with water molecules and silanol-terminate PDMS chains substituting for one another on boron crosslinks (Fig. \ref{fig:exchangeReaction}, where R = H). In the first step, a free water molecule substitutes for a PDMS chain on boron, resulting in an OH group on boron and a free silanol-terminated chain. In the second step, a free silanol-terminated PDMS chain substitutes for the OH group on boron, resulting in the PDMS chain on boron and a free water molecule. This water-catalyzed breaking and reforming of temporary crosslinks is analogous to the strand breaking and reforming of the TN model.

Carboxylic acids are far more effective than water at reducing $\tau$, probably because -COOH groups offer more opportunities for successful exchanges than -OH groups.

To compare the catalytic effectiveness of water and carboxylic acid, three versions of SB70-2.45 borosilicone (2.45 wt\% TMB in SFR 70) were prepared. One version was  simply vacuum-dried, the second version was vacuum-dried and then allowed to absorb $\sim$110 ppm ($\sim$0.011 wt\%) water from laboratory air, and the third version had 4 ppm (0.0004 wt\%) iso-stearic acid (ISA) added before vacuum-drying. The vacuum-dried version had $\tau=40.2$ s, the 110 ppm water version had $\tau=8.0$ s, and the 4 ppm ISA version had $\tau = 13.6$ s. It follows that ISA molecules reduce $\tau_{\textsc{tc}}$ approximately 250 times as effectively as water molecules.

Alcohols also reduce $\tau$, though less effectively than carboxylic acids or water. Moreover, alcohols tend to weaken or even liquefy borosilicones, an effect that makes alcohols useful for cleaning purposes. It may be that alcohol molecules substitute rapidly for PDMS chains on boron, but that the reverse substitution is considerably slower. If the resulting alkoxy groups on boron resist substitution, $\tau$ will decrease only slightly and the transient network will be weakened or even discontinuous.

Silanol groups themselves do not reduce $\tau$ significantly, probably because silanol-terminated PDMS chains rarely substitute for one another directly on boron crosslinks. As evidence for this limited exchange, consider SB40-4.50, a borosilicone with only 55\% of the boron crosslinks needed for stoichometric saturation (Table \ref{tab:OH40Viscosities}). Despite being at least 1.80 wt\% OH in the form of terminal silanol groups on PDMS chains, its $\tau$ can exceed 10 s. It follows that water molecules reduce $\tau_{\textsc{tc}}$ approximately 150 times as effectively as terminal silanol groups on PDMS.

The temperature dependence of a borosilicone's complex shear modulus $G^{*}(\omega)$ provides insight into its dynamics and energy barriers, so $G^{*}(\omega)$ was measured at several temperatures between 20 and 80 $^\circ$C. These temperature studies were conducted on two of the SB70-2.45 versions: the version containing $\sim$110ppm water and the version containing 4ppm ISA.

\begin{figure}
	\includegraphics{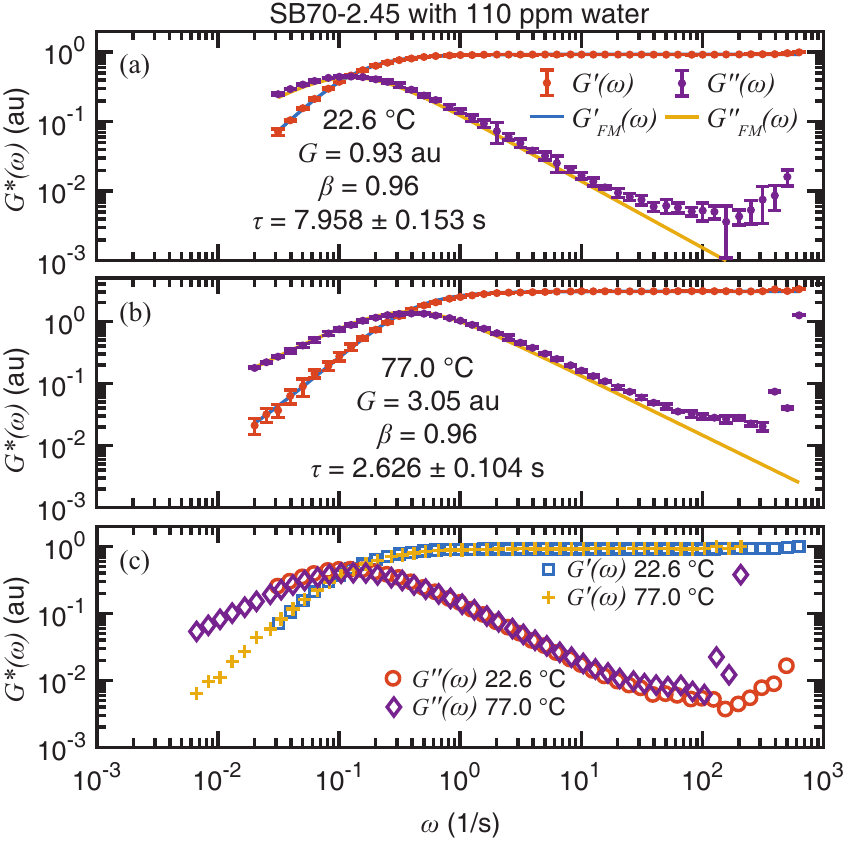}
	\caption{The complex shear modulus $G^{*}(\omega)$ for SB70-2.45 with 110ppm water. (a) was measured at 22.6 $^\circ$C and (b) at 77.0 $^\circ$C. (c) combines both sets of data, but with the higher temperature data set scaled vertically by the ratio of $G$ values ($\times 0.3050$) and horizontally by the ratio of $\tau$ values ($\times 0.3300$).\label{fig:RSWater}}
\end{figure}

\begin{figure}
	\includegraphics{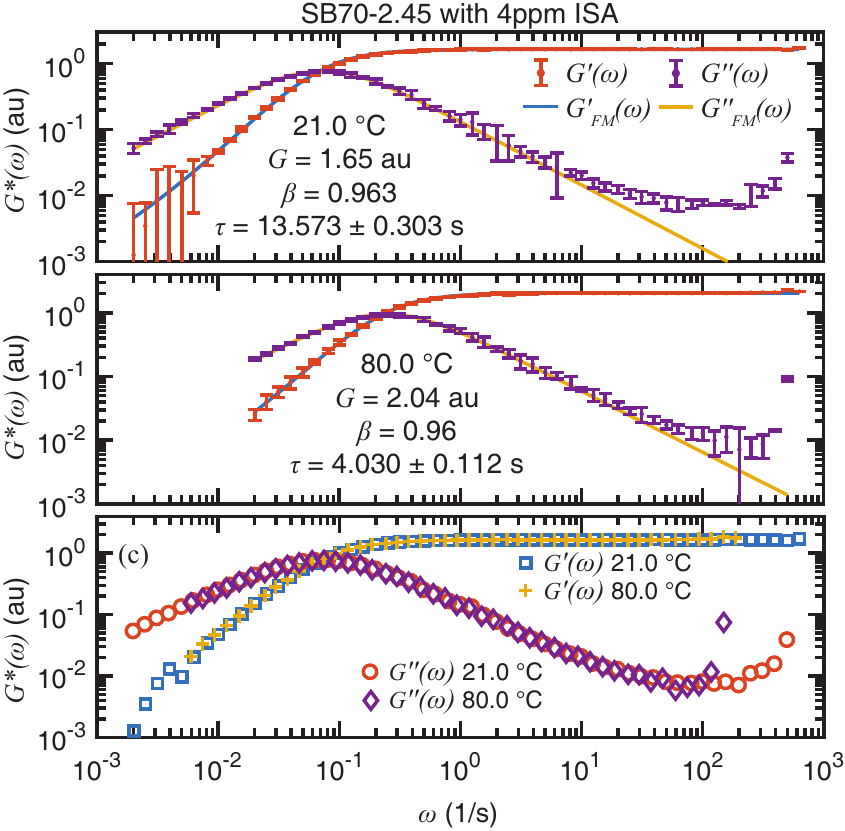}
	\caption{The complex shear modulus $G^{*}(\omega)$ for SB70-2.45 with 4ppm ISA. (a) was measured at 21.0 $^\circ$C and (b) at 80.0 $^\circ$C. (c) combines both sets of data, but with the higher temperature data set was normalized vertically ($\times 0.8074$) and shifted horizontally by the ratio of the $\tau$ values ($\times 0.2969$).\label{fig:RSISA}}
\end{figure}

\begin{figure}
	\includegraphics{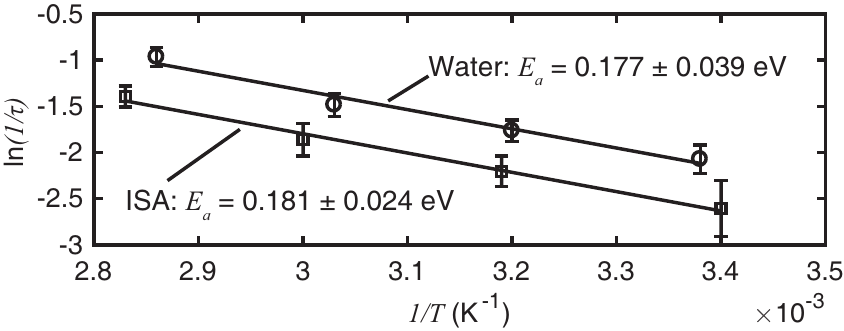}
	\caption{Arrhenium plots for SB70-2.45 containing 110ppm water ($\circ$) or 4ppm ISA ($\square$).\label{fig:figArrhenius}}
\end{figure}

The measured $G^{*}(\omega)$ curves were well-fit by $G_{\textsc{fm}}^{*}(\omega)$ (Figs. \ref{fig:RSWater}a,b and \ref{fig:RSISA}a,b), yielding $\beta$ values just slightly less than 1 and $\tau$ values that decrease modestly with increasing temperature. Arrhenius plots of $\ln(1/\tau)$ vs $1/T$ for both borosilicones are shown in Fig. \ref{fig:figArrhenius}. Assuming Eq. (\ref{eq:taus}), plots of $\ln(1/\tau_{\textsc{tc}})$ vs $1/T$ would look identical, except for a $\ln(1/a_{\textsc{b}})$ vertical shift.

The Arrhenius plot slopes for the 110ppm water and 4ppm ISA borosilicones give reaction energy barriers $E_{a}$ of $0.177\pm0.039$ eV ($4.07\pm0.90$ kcal/M) and $0.181\pm0.024$ eV ($4.16\pm0.54$ kcal/M), respectively. There is no significant difference between those energy barriers, so the difference in catalytic activities must be due to reaction cross sections: ISA has a much larger reaction cross section than water.

A material is said to be thermo-rheologically simple if, in the study of its linear viscoelastic behavior, a ``change in temperature is equivalent to a shift of the logarithmic time scale''\cite{schwarzl1952}. Figure \ref{fig:RSWater} shows $G^{*}(\omega)$ measurements of SB70-2.45 with $\sim$110ppm water taken at (a) 22.6 $^{\circ}$C and (b) 77.0 $^{\circ}$C. Both measurements appear in (c), with the 77.0 $^{\circ}$C scaled vertically by the ratio of $G$ values ($\times 0.3050$) and scaled horizontally by the ratio of $\tau$ values ($\times 0.3300$). Vertical scaling was necessary because of changes in sample shape between the two measurements. The excellent overlap of the two traces indicates that this borosilicone exhibits thermo-rheological simplicity. Increasing the temperature of the borosilicone reduces its characteristic time $\tau$ and thereby shifts the entire $G^{*}(\omega)$ to higher angular frequencies $\omega$.

Figure \ref{fig:RSISA} shows $G^{*}(\omega)$ measurements of SB70-2.45 with 4ppm ISA taken at (a) 21.0 $^{\circ}$C and (b) 80.0 $^{\circ}$C. Again, both measurements appear in (c), with the 80.0 $^{\circ}$C scaled vertically by ratio of $G$ values ($\times 0.8074$) and scaled horizontally by the ratio of $\tau$ values ($\times 0.2969$). This borosilicone also exhibits thermo-rheological simplicity. 
 
\section{Conclusions}

While boron atoms act as trifunctional crosslinkers that attach covalently to the ends of silicone polymer chains, the simple borosilicones they produce are not solids. Instead of gelling when its boron concentration exceeds the predicted gelation threshold, a simple borosilicone exhibits only a dramatic increase in shear viscosity. Its viscosity continues to increase as more boron is added, becoming extraordinarily large as the boron concentration approaches stoichiometric saturation. Regardless of boron concentration, a simple borosilicone never solidifies. The only plausible explanation for such observations is that boron crosslinks are temporary.

Boron crosslinks are temporary not because they are weak, but because they undergo easy substitution. The presence of -OH bearing molecules in the borosilicone allows for exchange reactions in which one -OH bearing molecular substitutes for another -OH bearing molecule on a boron crosslink. This exchange process is particularly efficient between carboxylic acids (R-COOH) and silanol-terminated silicones (OH-PDMS-OH). The relentless substitution of one attachment for another on each boron-silicone crosslink gives those crosslinks a mean lifetime that is typically measured in seconds or less. 

Because boron crosslinks are temporary, borosilicones are network liquids. Simple borosilicones, in which the silicone polymer chains are coupled only by temporary boron crosslinks, are excellent physical realizations of Green and Tobolsky's transient network model. Apart from some excess loss at large $\omega$, the observed moduli and viscosities of these simple borosilicones agree well with the predictions of the TN model and its Lodge Elastic Fluid. The Maxwell model, which defines a simple elastic fluid in terms of integer-order viscoelastic elements, shares those same predictions.

When permanent crosslinks are present in a borosilicone, it no longer behaves as a simple elastic fluid. Its time-dependences are slower than exponential and its storage and loss moduli decrease more slowly than $\omega^{2}$ and $\omega$, respectively, as $\omega\to 0$. Such observations cannot be modeled with finite assemblies of integer-order viscoelastic elements and the associated integer-order differential equations, so another approach is required.

The Random Assembly model, composed of infinitesimal viscous and elastic bodies  representing transient and non-transient couplings, provides a path forward. This model leads analytically and computationally to the spring-pot, a fractional-order viscoelastic element defined by fractional-order differential equations. The spring-pot's fractional order is $\beta$, which is also the volume concentration of viscous cuboids in the Random Assembly model.

When a spring is put in series with the spring-pot, to eliminate the spring-pot's divergence at $\omega\to\infty$ and recognize the finite compressibility of materials, the result is the Fractional Maxwell viscoelastic model. The observed moduli and viscosities of non-simple borosilicones agree well with the predictions of the FM model. The slower-than-exponential time-dependences observed experimentally correspond to Mittag-Leffler functions obtained from the FM model's fractional-order differential equations and the slow variations in storage and loss moduli observed as $\omega\to 0$ correspond to $\omega^{\beta}$ terms in the FM model solutions.

When the concentration of permanent crosslinks in a borosilicone exceeds the gelation threshold, its permanently-crosslinked subnetworks coalesce into a material-spanning network and the borosilicone becomes viscoelastic silicone rubber. The permanently-crosslinked network renders the VSR a network solid and gives it an equilibrium shape to which it returns when free of external influences. The VSR's temporary crosslinks, however, still form a liquid network, one so closely coupled to the solid network solid that it effectively ``piggybacks'' on the solid network.

The measured moduli of VSRs are well-described by the Fractional Zener model: a spring in parallel with the FM model. On the longest timescales and at the lowest frequencies, only the solid network's modulus is observed. On the shortest timescales and at the highest frequencies, the liquid network has no time to evolve and its full modulus is also observed. At intermediate timescales and frequencies, the dynamics of the liquid network are important and the VSR exhibits behaviors described by the fractional-order differential equations of the FM model. The fractional order $\beta$ depends primarily on the concentration of temporary crosslinks in the VSR and decreases as that concentration decreases.

In the end, borosilicones and VSRs are remarkably simple. On a molecular level, they are almost Tinkertoy\textsuperscript{\textregistered} systems: vast assemblies of elastic strands joined by temporary and permanent links or hubs. Their structural randomness and the statistical breaking of their temporary crosslinks further contribute to their simplicity. They permit simple structural modeling, including the transient network and random assembly models, and their behaviors can are beautifully characterized by few-element viscoelastic models, including the Maxwell, Fractional Maxwell, and Fractional Zener models.

These materials are not absolutely perfect, of course, so measurement and theory sometimes differ. What is consistent throughout this work, however, is the dominant role a single characteristic time $\tau$ plays in every measurement and every model of a borosilicone or VSR. In each case, $\tau$ derives from the mean lifetime of the material's temporary crosslinks and is the only significant timescale observed in the material.

The reciprocal of $\tau$ follows the Arrhenius equation, so the effect of a change in temperature is a change in $\tau$ and therefore a change in the material's only significant timescale. This behavior, in which a change in temperature causes only a change in timescale, is known as thermo-rheological simplicity. The borosilicones and VSRs studied in this work are thermo-rheologically simple.

\bibliographystyle{plainnat}
\bibliography{borosilicones}

\end{document}